\documentclass[11pt]{article}
\usepackage{amsfonts}
\usepackage{amssymb}
\usepackage{mathrsfs}
\usepackage{graphicx}
\usepackage{amsmath,amsthm,amssymb,amscd}
\usepackage{hyperref}

\newcommand{\ca}{{\cal A}}

\newcommand{\cn}{{\cal N}}

\newcommand{\cf}{{\cal F}}

\newcommand{\cs}{{\cal S}}

\def\eqa{\begin{eqnarray}}
\def\eqae{\end{eqnarray}}
\def\eq{\begin{equation}}
\def\eqe{\end{equation}}
\def\be{\begin{equation}}
\def\ee{\end{equation}}
\def\bea{\begin{eqnarray}}
\def\eea{\end{eqnarray}}
\def\ba{\begin{array}}
\def\ea{\end{array}}
\def\bd{\begin{displaymath}}
\def\ed{\end{displaymath}}

\def\Tr{{\rm Tr}}

\def\>{\rangle}
\def\<{\langle}
\def\a{\alpha}
\def\b{\beta}

\def\del{\delta}           
\def\e{\epsilon}           
  
\def\g{\gamma}
\def\h{\eta}

\def\l{\lambda}

\def\q{\theta}                  

\def\x{\xi}

\def\F{\Phi}

\def\pa{\partial}                              
\def\gt{generating function}

\newcommand{\lra}{\leftrightarrow}
\newcommand{\cyc}{{\rm cyc}}

\def\da{{\dot\alpha}}
\def\db{{\dot\beta}}

\def\tQ{\tilde{Q}}

\newcommand{\bh}{\overline{\eta}}

\newcommand{\reef}[1]{(\ref{#1})}

\numberwithin{equation}{section}

\begin{document}

\begin{titlepage}
\begin{flushright}
MCTP-11-08 \\
UCLA-11-TEP-001\\
\end{flushright}
\vspace{1cm}

\begin{center}
{\Large \bf On-shell superamplitudes in $\cn<4$ SYM}\\[1cm]
Henriette Elvang$^{a}$, Yu-tin Huang$^b$ and Cheng Peng$^a$\\
~\\[5mm]

$^a${\it Michigan Center for Theoretical Physics}\\
{\it Department of Physics, University of Michigan}\\
{\it Ann Arbor, MI 48109, USA}\\
~\\

$^b${\it Department of Physics and Astronomy}\\
 {\it UCLA, Los Angeles}\\
 {\it CA 90095, USA}

~\\
~\\
\today
~\\
~\\

\end{center}

\begin{center}
Abstract\\
\end{center}

We present an on-shell formalism for superamplitudes of pure 
$\cn<4$ super Yang-Mills theory. Two superfields, $\Phi$ and $\Phi^\dagger$,  are required to describe the two CPT conjugate supermultiplets.    Simple truncation prescriptions allow us to derive explicit tree-level MHV and NMHV  superamplitudes with $\cn$-fold SUSY. Any $\cn\!=\!0,1,2$ 
tree superamplitudes have large-$z$ falloffs under super-BCFW shifts, except under $[\Phi,\Phi^\dagger\>$-shifts. We show 
that this `bad' shift is responsible for the bubble contributions to 1-loop amplitudes in  \mbox{$\cn\!=\!0,1,2$ SYM}. We evaluate the MHV bubble coefficients in a manifestly supersymmetric form and demonstrate for the case of four external particles that the sum of bubble coefficients
is equal to minus the tree superamplitude times the 1-loop beta-function coefficient. The connection to the beta-function is expected since only 
bubble integrals capture UV divergences; we discuss briefly 
how the minus sign arises from UV and IR divergences in dimensional regularization.

Other applications of the on-shell formalism include a solution to the N$^K$MHV  $\cn\!=\!1$ SUSY Ward identities and a clear description of the connection between 6d superamplitudes and the 4d ones for both $\cn\!=\!4$ and $\cn\!=\!2$ SYM. We outline extensions to $\cn\!<\!8$ supergravity.

\end{titlepage}

\tableofcontents
\newpage

\section{Introduction and summary of results}\label{introduction}

The study of on-shell scattering amplitudes has in recent years revealed many surprising structures and completely new ways to evaluate amplitudes at both tree and loop level. Particularly remarkable results have been found in the planar sector of massless $\cn=4$ SYM theory. It is obviously of considerable practical and theoretical interest to generalize the results of this very special integrable sector to theories 
with less symmetry. In this paper we take a step towards this goal by studying basic properties of scattering amplitudes in pure $\cn=1$ and $\cn=2$ SYM.

A cornerstone in the recent developments in $\cn=4$ SYM  has been the on-shell superfield formalism which encodes amplitudes related by supersymmetry into \emph{superamplitudes} \cite{Nair:1988bq,Bianchi:2008pu,Drummond:2008vq,Elvang:2008na,Elvang:2008vz}. 
In this formalism, the CPT self-conjugate supermultiplet of on-shell states is collected into a superfield, or \emph{superwavefunction},
\begin{equation}\label{n4Om}
    \Omega_i=G_i^++\h_{ia}\,\l_i^{a}- \frac{1}{2!}\h_{ia}\h_b\, S_i^{ab}
    -\frac{1}{3!}\h_{ia}\h_{ib}\h_{ic}  \,{\l}_i^{abc}+\h_{i1}\h_{i2}\h_{i3}\h_{i4}\, G_i^- \,.
\end{equation}
The Grassmann variables $\eta_{ia}$ are labeled by particle number $i$ and $SU(4)$ R-symmetry index $a=1,2,3,4$, and the components are on-shell states --- gluons $G$, gluinos $\lambda$, and scalars $S$ ---
with momentum $p_i$.  The superamplitude $\ca_n(\Omega_1,\Omega_2,\dots,\Omega_n)$ is a polynomial in the $\eta_{ia}$-variables and each coefficient is an on-shell scattering amplitude. To project out a particular scattering amplitude from $\ca_n$ we act with the unique set of Grassmann derivatives that project out the desired set of external states from the superwavefunctions $\Omega_i$. 
Thus superamplitudes are generating functions for the component amplitudes. 

\vspace{2mm}
The superamplitude formalism has been 
central in many recent developments:
\begin{itemize}
 \item {\bf Super-BCFW recursion relations.}
The familiar BCFW shift \cite{Britto:2004ap,Britto:2005fq} can be
accompanied by a shift of the Grassmann variables $\eta_{ia}$ such
that the supermomenta  $\tilde{Q}_a$ and $\tilde{Q}^a$ are
unshifted \cite{ArkaniHamed:2008gz,Brandhuber:2008pf}.  The resulting super-BCFW recursion relations are valid
for all $\cn=4$ SYM superamplitudes with $n \ge 4$
\cite{ArkaniHamed:2008gz}. They have been crucial for multiple
recent developments, including
\cite{Drummond:2008cr,ArkaniHamed:2009si,ArkaniHamed:2009dn,
ArkaniHamed:2010kv,ArkaniHamed:2010gg,Drummond:2009fd,Mason:2009sa}.

 \item {\bf Dual superconformal symmetry.}
 Dual conformal symmetry acts as ordinary conformal transformations
 on the momentum ``region variables'' $x_i$, defined by $p_i = x_i -
 x_{i+1}$. For example $x_i \to x_i/x_i^2$ under dual conformal inversion.  
 Split-helicity gluon amplitudes, such as $A_{6}(--++++)$, transform covariantly under dual conformal symmetry, but non-split amplitudes such as $A_{6}(-+-+++)$ do not have decent transformation properties. It is only when the component amplitudes are collected into super\-amplitudes  $\mathcal{A}_n$
 that the dual superconformal symmetry reveals itself
 \cite{Drummond:2008vq,Brandhuber:2008pf,Drummond:2008cr}. 
 Dual superconformal symmetry is a property of \emph{planar} amplitudes in the strong coupling regime \cite{Alday:2007hr} and perturbatively at tree-level \cite{Brandhuber:2008pf,Drummond:2008cr} as well as loop-level \cite{Drummond:2006rz,Drummond:2008vq,Brandhuber:2009xz,Elvang:2009ya,Brandhuber:2009kh,ArkaniHamed:2010kv}. 
The dual and ordinary superconformal algebra form the first two
levels of a Yangian \cite{Drummond:2009fd}, and constructing 
Yangian invariants has been a central guiding principle for the recently developed loop-level recursion relations \cite{ArkaniHamed:2010kv}. On-shell superfield formulations were also needed for dual conformal symmetry of planar amplitudes of maximally supersymmetric Yang-Mills theory in six \cite{6loopDamplitudeYT,Dennen:2010dh} and ten dimension \cite{CaronHuot:2010rj}.

 \item {\bf Efficient evaluation of intermediate state sums.}
The superamplitude formulation makes it possible to efficiently and systematically perform intermediate state sums \cite{Bianchi:2008pu,Elvang:2008na} in (unitarity cuts of) loop amplitudes. 
These super-sums are carried out as Grassmann integrals over all $\eta_{\ell a}$'s associated with the internal lines $\ell$ of loop diagram \cite{Bianchi:2008pu}. This evaluation method was used in certain cuts needed in the recent 4-loop $\cn=4$ SYM calculation \cite{Bern:2010tq} and are valid for both planar and non-planer contributions. A detailed analysis of super-sums together with a diagrammatic representation was given in \cite{Bern:2009xq}.

 \item {\bf Solution to the SUSY Ward identities.}
On-shell Ward identities of supersymmetry enforce linear
relationships among the amplitudes in each N$^K$MHV sector. These are trivially solved at MHV level, but for NMHV and beyond the coupled linear systems are nontrivial and appear quite intractable. However, the SUSY Ward identities can be formulated as the requirement that the SUSY charges annihilate the superamplitudes, and in this language the problem has a simple solution that expresses the superamplitude $\ca_n$ as a sum of manifestly SUSY and R-symmetry invariant Grassmann polynomials \cite{Elvang:2009wd,Elvang:2010xn}.

 \item {\bf Supergravity.}
 The  UV structure of perturbative $\cn=8$ supergravity 
 can be investigated via a characterization of available candidate
counterterms. On-shell superamplitude techniques have recently been used to examine the
matrix elements produced by putative counterterm operators 
\cite{Elvang:2010jv}. Analysis of the combined requirements  of $\cn=8$ SUSY, 
full $SU(8)$ R-symmetry, and the low-energy theorems of the spontaneously broken $E_{7(7)}$-symmetry
(see
\cite{Bianchi:2008pu,ArkaniHamed:2008gz,Elvang:2010kc,Beisert:2010jx,Kallosh:2008rr,Brodel:2009hu,
Bossard:2010dq,Bossard:2010bd}) shows that no divergences are expected in 4d amplitudes until 7-loop order \cite{Elvang:2010jv,Elvang:2010kc,Beisert:2010jx} (see
also \cite{Drummond:2003ex,Drummond:2010fp}).  

\end{itemize}

Superamplitudes and on-shell superspace techniques of $\cn=4$ SYM and $\cn=8$ supergravity were needed in all these examples. The purpose of the present paper is to develop on-shell superfield formalisms for pure $\cn<4$ SYM theory (and we will briefly comment on $\cn<8$ supergravity). While the spectrum of $\cn=4$ SYM is
CPT self-conjugate, this is not the case for SYM theory with less supersymmetry. Thus two superfields are needed to encode the spectrum of pure\footnote{Unless otherwise stated, we use  $\cn<4$ SYM to refer to \emph{pure}  $\cn<4$ SYM.} $\cn<4$ SYM theory: one superfield for the `positive helicity sector' and one for the `negative helicity sector'. For example, for $\cn=1$ SYM we use
\bea
  \label{n1intro}
  \Phi = G^+ +\h_{i}\,\l^+ \, ,~~~~~
  \Phi^\dagger = G^- +\bar{\h}_{i}\,\l^- \, .
\eea
Note that $\Phi$ is simply the truncation $\eta_{i2},\eta_{i3},\eta_{i4} \to 0$ of the $\cn=4$ superfield \reef{n4Om} while $\Phi^\dagger$ can be obtained from  \reef{n4Om} by carrying out a Fourier transformation of the Grassmann variables and then taking $\bar\eta_{i2},\bar\eta_{i3},\bar\eta_{i4} \to 0$. Clearly this procedure  can be exploited to systematically truncate $\cn=4$ SYM superamplitudes at \emph{tree level} to $\cn=1$ SYM, and more generally to $\cn=0,1,2,3$ SYM. This works because 
$\cn=0,1,2$ SYM form closed subsectors of the $\cn=4$ theory at tree level. The $\cn=3$ formulation provides a non-chiral\footnote{We use ``chiral'' to denote on-shell superspace with only $\eta$-Grassmann variables; thus ``non-chiral'' means that both $\eta$ and $\bar\eta$ are used.} but otherwise equivalent on-shell superspace formulation of the $\cn=4$ theory.

N$^K$MHV superamplitudes in $\cn<4$ SYM involve 
$(K\!+\!2)$ $\Phi^\dagger$ superwavefunctions and \mbox{$(n\!-\!K\!-\!2)$}  $\Phi$'s.  Since the amplitudes are color-ordered and $\cn<4$  
 SUSY does not mix the states of the `positive' and `negative  \mbox{helicity}  sectors', there are now superamplitudes 
for each arrangement of the 
$\Phi^\dagger$ and $\Phi$ states. For example, the MHV superamplitudes $\ca_{n;12}(\Phi_1^\dagger \Phi_2^\dagger \Phi_3 \dots \Phi_n)$ and $\ca_{n;13}(\Phi_1^\dagger \Phi_2 \Phi_3^\dagger \Phi_4 \dots \Phi_n)$ are distinct. The formalism is discussed in {\bf section \ref{sec:MHVgt}}.

The non-chiral $\Phi$-$\Phi^\dagger$
formulation \reef{n1intro} of $\cn<4$ SYM turns out to be somewhat impractical for explicit calculations, and it is convenient to replace $\Phi^\dagger$ by its Fourier transform $\Psi$. We introduce the chiral $\Phi$-$\Psi$ formalism in {\bf section \ref{s:PhiPsi}} and  apply it in subsequent sections. 
In this formalism, the tree-level MHV amplitudes in pure SYM with $0 \le \cn \le 4$ can be written compactly as (see also \cite{Bern:2009xq})
\eq
  \label{MHVgenNintro}
  \cf_{n,ij}^{\mathcal{N}}
 ~=~
  (-1)^{\frac{1}{2}\cn(\cn-1)} ~\frac{\langle ij\rangle^{4-\cn}~\delta^{(2\cn)} \big(\sum |k\> \eta_k \big)}{\langle12\rangle\langle 23\rangle \cdot\cdot \langle n\,1\rangle}\,,
 \eqe
where the Grassmann delta-function expresses conservation of the $\cn$ supermomenta; the standard momentum delta-function is implicit. We derive similar explicit
$0 \le \cn \le 4$ formulas for the NMHV superamplitudes and discuss the general truncation procedure beyond NMHV. The resulting formalism should be straightforward to incorporate into numerical programs such as the Mathematica packages presented recently in \cite{Dixon:2010ik,Bourjaily:2010wh}.

\vspace{2mm}
As applications of the $\cn<4$ superamplitude formalism we study:\footnote{Dual superconformal symmetry is not on the list of properties we explore in $\cn<4$ SYM simply because in general it is not a property of the amplitudes.}
\begin{itemize}

\item[$\triangleright $] {\bf Section \ref{s:bcfw}: Super-BCFW recursion relations.} We formulate super-BCFW shift in  $\cn < 4$ SYM, and show that the large-$z$ behavior can be derived by a simple Grassmann integral argument. The tree-level superamplitudes in $\cn=4$ SYM have large-$z$ falloff under any super-BCFW shift. In $\cn<4$ SYM, the shifts $[\Psi,\Psi\>$, $[\Psi,\Phi\>$, $[\Phi,\Phi\>$ give similar large-$z$ falloffs and the associated super-BCFW recursion relations are therefore valid. However, under a $[\Phi,\Psi\>$-shift, the $\cn$-fold superamplitudes  behave
as $z^{3-\cn}$  (adjacent; $z^{2-\cn}$ for non-adjacent) for large $z$; we study the consequences at loop level.

\item[$\triangleright $] {\bf Section \ref{s:1loop}. 
Structure of 1-loop amplitudes: supersums, bubble contributions and UV \& IR divergences.}  
1-loop amplitudes of SYM can be reconstructed completely from their unitarity cuts, and an explicit expansion involves scalar box, triangle and bubbles integrals \cite{GenUni,Bern:1994zx}.  Bubble cuts involve a product of two tree amplitudes. Following the work of Forde \cite{Forde:2007mi}, Arkani-Hamed, Cachazo and Kaplan 
 \cite{ArkaniHamed:2008gz} showed that the bubble coefficients are non-vanishing when this product has an $O(1)$-term for large-$z$ under a (super-)BCFW-shift of the loop-momentum. 
 The fact that all super-shifts give large-$z$ falloff in $\cn=4$ SYM implies that bubbles are absent.

The result that $\cn=0,1,2$ SYM tree superamplitudes do not falloff for large $z$ under  $[\Phi,\Psi\>$-shifts allow us to identity which bubble coefficients are non-vanishing without explicit calculation. We then proceed to evaluate these coefficients  using the results for the tree superamplitudes to  compute $\cn=1,2$ super-sums. For 4-point amplitudes we carry out the $d$LIPS-integrals and demonstrate for  that the sum of bubble coefficients
equals  $- \beta_0 {\cal A}_4^\text{tree}$ with $\beta_0$ 
the 1-loop $\beta$-function coefficient. 
The equivalent result was obtained in \cite{ArkaniHamed:2008gz} for pure $\cn=0$ YM; we find that performing the intermediate state sum before evaluating the $d$LIPS integral yields the result more directly. The connection to the 1-loop $\beta$-function is natural since only the bubbles capture UV divergences.

Our work on the 1-loop structure of Yang-Mills amplitudes overlaps with the interesting work of Lal and Raju \cite{Lal:2009gn}. They used an on-shell superspace formalism to analyze conditions for the absence of triangle and bubble contributions to the 1-loop amplitude in gauge theories. In contrast, we use the on-shell formalism to find explicit results for the bubbles in pure $\cn=1,2$ SYM.

\item[$\triangleright $] {\bf Section \ref{sec:SWI}: Solution to the SUSY
Ward identities in $\cn=1$ SYM.}  The SUSY Ward identities in
$\cn=1$ SYM are even simpler to solve than in $\cn=4$ SYM. A total
of $\binom{n-4}{K}$ algebraically independent basis amplitudes
determine the N$^K$MHV superamplitudes for  each arrangement of
external states $\Phi$ and $\Psi$. For $n=6$ and NMHV ($K\!=\!1$)
the counting of 2 basis-amplitudes agrees with the only previous
solution \cite{Grisaru:1977px,Bianchi:2008pu,Brodel:2009hu} for
$\cn=1$ SYM.

\end{itemize}

Amplitudes in $D\ne 4$ have been explored in various works  \cite{6Dspinor1,6Dspinor2,6Dspinor3,6D20,CSetc}. 
The 6d maximally SYM theory has $\cn=(1,1)$ supersymmetry, and restricting its momenta to a 4d subspace gives 
massless $\cn=4$ SYM. The on-shell superspace formalism for 6d superamplitudes is non-chiral and
yields upon reduction to 4d a non-chiral representation of the $\cn=4$ SYM superamplitudes \cite{Hatsuda:2008pm}. Following  \cite{Hatsuda:2008pm}, we present  in {\bf section \ref{sec:6d}} the precise map to convert the 4d reduction of the 6d superamplitudes to the familiar chiral form and discuss how the 4d N$^K$MHV 
helicity sectors emerge. We show how to truncate the 6d $\cn=(1,1)$ SYM tree-level superamplitudes to $\cn=(1,0)$ SYM, which upon reduction to 4d  yields the tree-level superamplitudes of $\cn=2$ SYM.

We end our story with two short sections: in {\bf section \ref{s:SG}} we outline the superfield formalism for superamplitudes of $\cn<8$ supergravity, and in the Outlook, {\bf section \ref{sec:disc}},  we briefly discuss the coupling of matter multiplets to the $\cn=1,2$ SYM theories.  A few technical details are stowed
away into two appendices.


\section{On-shell formalism for pure SYM: $\Phi$-$\Phi^\dagger$ formalism}\label{sec:MHVgt}
To set the stage for $\cn<4$ SYM, we begin with a brief review of the relevant on-shell framework in $\cn=4$ SYM.

\subsection{On-shell superfields and MHV superamplitudes in $\cn=4$ SYM}

The on-shell supermultiplet of $\cn=4$ SYM consists of 16 massless  particles:
\bea
\text{two gluons $G^\pm$,~~ four gluinos pairs $\l^a$ and $\l^{abc}$,~~ and six scalars $S^{ab}$}\,.
\eea
 The indices $a,b,\dots = 1,2,3,4$ are $SU(4)$ R-symmetry labels. The helicity $h = \pm1, \pm\tfrac{1}{2},0$ states transform as $r$-index anti-symmetric representations of $SU(4)$ with $r=2-2h$.
We collect the 16 states into an $\cn=4$ on-shell chiral superfield
\begin{equation}\label{n4Phi}
    \Omega=G^++\h_a\l^{a}- \frac{1}{2!}\h_a\h_b S^{ab}
    -\frac{1}{3!}\h_a\h_b\h_c  {\l}^{abc}+\h_1\h_2\h_3\h_4\, G^- \,,
\end{equation}
where the four $\h_a$'s are Grassmann variables labeled by the $SU(4)$ index $a=1,2,3,4$. These variables were first introduced by Ferber \cite{Ferber1977qx} as the superpartners to the bosonic twistor variables.  
The relative signs are chosen such that the Grassmann differential operators
\bea
\label{n4ops}
\cn=4~\text{SYM:}~~~~
  \begin{array}{c|c|c|c|c|c}
  \text{particle} & G^+ & \lambda^{a} & S^{ab}
  & \lambda^{abc} &G^{1234} \\[1mm]
  \hline
  \raisebox{-1.5mm}{operator}  &
  \raisebox{-1.5mm}{1}
  & \raisebox{-1.5mm}{$\partial_{i}^a$}
  & \raisebox{-1.5mm}{$\partial_{i}^a\partial_{i}^b$}
  & \raisebox{-1.5mm}{$\partial_{i}^a\partial_{i}^b\partial_{i}^c$}
  & \raisebox{-1.5mm}{$\partial_{i}^1\partial_{i}^2\partial_{i}^3\partial_{i}^4$}
\end{array}
\eea
exactly select the associated state from $\Omega$.

All 16 states of the multiplet are related by supersymmetry. In the
 on-shell formalism the supercharges are
\begin{equation}
\label{qtq}
q^a \equiv q^{a\,\a} \,\e_{\a} = - [p\,\e]\,\frac{\pa}{\pa\h_a} \,,
~~\qquad
\tilde{q}_a \equiv  \tilde{\e}_{\da}\, \tilde{q}_{a}^{\da} = \<\e \,p\>\,\h_a \,,
\end{equation}
with $|p\>$ and $|p]$ the spinors associated with the null momentum $p$ of the particle. $\e$ is an arbitrary Grassmann spinor. The supercharges satisfy the
anticommutation relation
\begin{equation}\label{qcom}
    \{q^{a\, \a},\tilde{q}^{\db}_b\}= \delta^a_{b}\,|p\>^{\db}[p|^{\a}=\delta^a_{b}\, p^{\dot{\b}\a}
\end{equation}
of the Poincare supersymmetry algebra.

The supercharges \reef{qtq}  act on the spectrum by shifting states right or left in $\Omega$. For example, if we compare $\Omega$ with
\bea
  \tilde{q}_1 \, \Omega
    &=&
    -\Big(
    \eta_1   \<\e p\> G^+
    - \eta_1 \eta_a  \<\e p\> \l^1
    - \frac{1}{2} \eta_1 \eta_a \eta_b \,  \<\e p\> S^{ab}
    +   \eta_1 \eta_2 \eta_3 \eta_4  \<\e p\>\l^{234}
  \Big)
  \label{tq1Phi}
\eea
order by order in $\eta$'s to extract the action of $\tilde{q}_1$ on the individual states, we find\footnote{We drop an overall sign in \reef{tq1Phi}.}
\bea
\nonumber
&&\big[\tilde{q}_a,G^+\big] =0\, ,~~~~~~
\big[\tilde{q}_a,\l^b\big] = \<\epsilon\, p\>\,\delta^b_a\,G^+\, ,~~~~~~
\big[\tilde{q}_a,S^{bc}\big]
=\<\epsilon\,p\> \, 2! \, \delta^{[b}_a\,\l^{\raisebox{0.6mm}{\scriptsize$c]$}}\, ,
\\[2mm]
&&
\big[\tilde{q}_a,\l^{bcd}\big] =\<\epsilon\, p\>\, 3!\, \delta_a^{[b} S^{\raisebox{0.6mm}{\scriptsize$cd]$}}\, ,~~~~~~
\big[\tilde{q}_a,G^{bcde}\big] = \<\epsilon\,p\>\, 4!\, \delta_a^{[b} \l^{\raisebox{0.6mm}{\scriptsize$cde]$}} \, .
\label{tqaction}
\eea
Similar relations are found for $q^a$.
Note that the action of $\tilde{q}_a$ on operators in \reef{n4ops} is identical to \reef{tqaction}. However, in this chiral representation, the $q^a$'s commute with all the Grassmann differential operators in \reef{n4ops}.

Superfields $\Omega$ can be regarded as \emph{superwavefunctions} for the external lines of superamplitudes $\ca_n(\Omega_1,\Omega_2,\dots,\Omega_n)$. The N$^K$MHV superamplitudes of $\cn=4$ SYM are degree $4(K+2)$ polynomials in the $n$ sets of Grassmann variables $\eta_{ia}$. The individual amplitudes are coefficients of this polynomial. One extracts an amplitude by applying the operators \reef{n4ops} to $\ca_n$ to project out each of the desired states from the superwavefunctions $\Omega_i$.

The tree-level MHV superamplitude \cite{Nair:1988bq} is simply given by
\begin{equation}\label{N4MHV}
{\cal A}^{\text{MHV}}_{n}
=
  \frac{\delta^{(8)}(\tilde{Q})}{\<12\>\<23\>\ldots\<n1\>}\,,
\end{equation}
where the Grassmann delta-function is defined as
\bea
  \delta^{(8)}(\tilde{Q}) \equiv
  \frac{1}{2^4}
  \prod\limits_{a=1}^4
  \sum_{i,j=1}^n\langle ij\rangle\eta_{ia}\eta_{ja}\,.
\eea
The sums of supercharges \reef{qtq}
\bea
 \tilde{Q}^{\dot{\a}}_a = \sum_{i=1}^n |i\>^{\dot{\a}}\, \eta_{ia}\,,\qquad
 Q^a_\a = \sum_{i=1}^n |i]_\a\, \frac{\partial}{\partial\eta_{ia}}
\eea
both annihilate $\delta^{(8)}(\tilde{Q})$, so the MHV superamplitude
${\cal A}^{\text{MHV}}_{n}$ is manifestly supersymmetric.
There are known tree-level expressions for all N$^K$MHV amplitudes of $\cn=4$ SYM \cite{Drummond:2008cr}. In this section we only consider MHV superamplitudes, but we go beyond MHV in section \ref{s:PhiPsi}.

We have used a superfield $\Omega$ chiral in $\eta_a$ to encode the states of $\cn=4$ SYM. The conjugate superfield $\Omega^\dagger$ encodes exactly the same
information as $\Omega$, since the $\cn=4$ SYM multiplet is CPT self-conjugate. The equivalence of the fields are easily seen by a Grassmann Fourier transformation; indeed one finds
\bea
   \nonumber
    \Omega^{\dagger}
    &=& \int d\eta_1 d\eta_2 d\eta_3 d\eta_4
    ~e^{\eta_a \overline{\eta}^a}~\Omega\\
    &=&
    G^- + \frac{1}{3!}\e_{abcd} \bh^a \l^{bcd}
    +\frac{1}{2!}\frac{1}{2!}\e_{abcd}\bh^a\bh^b \,S^{cd}
    +\frac{1}{3!}\e_{abcd}\bh^a\bh^b\bh^c \, \l^{d}
    +\bh^1\bh^2\bh^3\bh^4 \,G^+ \,.~~~~
    \label{n4Phid}
\eea
Comparing with the directly conjugated field, we have identified $(G^+)^\dagger = G^-$, including the anti-self-conjugacy condition
$\bar{S}_{ab} = -\tfrac{1}{2!}\e_{abcd}S^{cd}$ for the scalars.

Since the two wavefunctions $\Omega$ and $\Omega^\dagger$ encode the exact same information, we are free to use either formulation in the superamplitudes. This will be useful in the following.

\subsection{$\cn=1$ SYM on-shell superfields $\Phi$ and $\Phi^\dagger$ and MHV superamplitude}\label{n1superfield}

The $\cn= 1$ SYM supermultiplet consists of a gluon,
$G^+$, with helicity $+1$ and a gluino, $\l^+$, with helicity
$+1/2$, and in addition the CPT conjugate gluon $G^-$ and gluino $\l^-$ with negative helicities. Classically, pure $\cn=1$ SYM theory has a $U(1)_R$ global symmetry, under which the particles have the R-charges
\bea
\begin{array}{c|c|c|c|c}
  & G^{+} & \l^+& \l^- & G^-
\\\hline
\text{R-charge} & 0& 1&-1&0
\end{array}
\eea
It is natural to encode the $\cn=1$ states into two conjugate on-shell superfields
\begin{equation}\label{superwf}
    \F=G^++\h\,\l^+ \,, \qquad
    \F^{\dagger}=G^-+\bh\,\l^- \,.
\end{equation}
The $\cn=1$ theory forms a closed subsector of the $\cn=4$ theory, and  the $\cn=1$ wavefunctions $\F$ and  $\F^\dagger$ in \reef{superwf} can be obtained from the $\cn=4$ superfields \reef{n4Phi} and \reef{n4Phid} by a truncation
\bea
  \label{trunc1}
 \eta_{2,3,4} \to 0\,, \qquad
 \overline{\eta}_{2,3,4} \to 0
\eea
with the identification $\lambda^+ = \l^1$ and $\l^- = \lambda^{234}$.

Let us now use this to obtain the MHV tree superamplitudes in $\cn=1$ SYM.
If we perform the truncation \reef{trunc1} directly on the $\cn=4$ MHV superamplitude \reef{N4MHV}, it clearly vanishes. This is not surprising because  it would correspond to an amplitude with external states only from the positive helicity sector of $\cn=1$ SYM, and this is forbidden by supersymmetry.

We recall that the MHV sector in $\cn=1$ SYM consists of $n$-point amplitudes with two states from the negative helicity sector $\F^\dagger$ and $n-2$ from the positive helicity sector $\F$. It is therefore natural that $\cn=1$ SYM superamplitudes in the MHV sector take the form
\begin{equation}\label{Superamplitudesgeneralmhv}
    \ca^{\cn=1}_{n,ij}=
    \ca^{\cn=1}_n(\F_1\F_2 \ldots \F^{\dagger}_{i}\, \F_{i+1}
    \ldots \F^{\dagger}_{j}\,\F_{j+1}\ldots\F_n)\,.
\end{equation}
The subscript $ij$ on $\ca^{\cn=1}_{n,ij}$ indicate the states in the $\F^{\dagger}$ sector.

The equivalence between the description of the $\cn=4$ supermultiplet in the $\Omega$ or $\Omega^\dagger$ superfields can now be exploited to obtain the $\cn=1$ SYM MHV superamplitudes $\ca^{\cn=1}_{n,ij}$ in two easy steps. The first step is to perform a Grassmann Fourier transform of the $\eta$-variables of lines $i$ and $j$ in
the $\cn=4$ superamplitude \reef{N4MHV}. This converts $\Phi_i$ and $\Phi_j$ to $\Phi_i^\dagger$ and $\Phi_j^\dagger$, and thus yields the equally valid $\cn=4$ SYM MHV superamplitude\footnote{We define $d^4\eta_i \equiv \prod_{a=1}^4 d{\eta_{ia}}$.}
\bea
\nonumber
&&
\hspace{-1.1cm}
{\cal A}^{\cn=4,\text{MHV}}_{n,ij}
  ( \ldots \F^{\dagger}_{i}
    \ldots \F^{\dagger}_{j}\dots)
    =\int d^4\eta_i\, d^4\eta_j~
e^{\eta_{i b} \overline{\eta}_{i }^b}~
e^{\eta_{jc}\overline{\eta}_{j }^c}~
{\cal A}^{\text{MHV}}_{n}
  (\F_1\F_2 \ldots\F_n)
    \\
&&
  =\frac{1}{\<12\>\<23\>\ldots\<n1\>}\prod\limits_{a=1}^4
  \Big(\langle ij \rangle
  +\langle i k \rangle\, \overline{\eta}_{j}^a\, \eta_{ka}
  -\langle j k \rangle\, \overline{\eta}_{i}^a\, \eta_{ka}
  -\frac{1}{2}\langle  k l\rangle  \,
     \overline{\eta}_{i}^a\,\overline{\eta}_{j}^a\, \eta_{ka}\,\eta_{la}
     \Big)\,.~~~~
     \label{MHVstep1}
\eea
There is an implicit sum over repeated indices $k,l=1,2,\dots, n$ with $k,l\ne i,j$.
The second step is to apply the truncation \reef{trunc1} to \reef{MHVstep1} to find the $\cn=1$ MHV superamplitudes:
\begin{equation}
\text{MHV}:~~
{\cal A}^{\cn=1}_{n,ij}~=~
\frac{\langle ij \rangle^3}{\<12\>\<23\>\ldots\<n1\>}   \Big(\langle ij \rangle
  +\langle i k \rangle\, \overline{\eta}_{j}\, \eta_{k}
  -\langle j k \rangle\, \overline{\eta}_{i}\, \eta_{k}
  -\frac{1}{2}\langle  k l\rangle  \,
     \overline{\eta}_{i}\,\overline{\eta}_{j}\, \eta_{k}\,\eta_{l}
     \Big)
  \,.\label{F1MHV}
\end{equation}
The choice  of $\Phi^\dagger$ states $i$ and $j$ necessarily breaks the cyclic symmetry of the original $\cn=4$ superamplitude.

\begin{table}[t]
\centering{
   \begin{tabular}{c|c|c}
  \multicolumn{3}{c}{$\cn=1$ SYM}
  \\[2mm]
  \hline
  particle & operator $\Phi$-$\Phi^\dagger$
  & operator $\Phi$-$\Psi$   \\
  \hline\hline
  \raisebox{-1.5mm}{$G^+$}
  & \raisebox{-1.5mm}{1}  & \raisebox{-1.5mm}{1}   \\[2.5mm]
  \hline
  \raisebox{-1.5mm}{$\lambda^+$}
  & \raisebox{-1.5mm}{$\partial_i$}
  & \raisebox{-1.5mm}{$\partial_i$} \\[2.5mm]
  \hline
  \raisebox{-1.5mm}{$\lambda^-$}
  & \raisebox{-1.5mm}{$\overline\partial_i$}
  & \raisebox{-1.5mm}{1}
  \\[2.5mm]
  \hline
  \raisebox{-1.5mm}{$G^-$} & \raisebox{-1.5mm}{\={1}}
   & \raisebox{-1.5mm}{$\partial_i$}
  \\[2.5mm]\hline
\end{tabular}
        }
\caption{
\small
Map from states to Grassmann derivatives for $\cn=1$ SYM.
We present two different formalisms, one with on-shell superfields $\Phi$-$\Phi^\dagger$ and conjugate Grassmann variables $\eta_{ia}$ and $\bar{\eta}_{i}^a$ (section \ref{n1superfield}), and the other with superfields $\Phi$-$\Psi$  chiral in Grassmann variables $\eta_{ia}$ (section \ref{s:PhiPsi}).}
\label{mapping}
\end{table}

Explicit amplitudes are projected out by acting on ${\cal A}^{\cn=1}_{n,ij}$ with Grassmann derivatives that select the requested external states from the superfields \reef{superwf} \emph{and} then set any remaining $\eta$-variables to zero. (Equivalently, we can convert the Grassmann differentiations to integrals.)
The map between states and Grassmann derivative operators is summarized in table \ref{mapping}. We list three simple examples:
\begin{eqnarray}
\nonumber
\big\langle--++++\big\rangle &=&
{\cal A}_{6,12}\big|_{\eta \to 0} ~=~
\frac{\langle12\rangle^4}{\<12\>\<23\>\ldots\<61\>}\,,
\\
\label{testresult}
  \big\langle-\lambda^-\lambda^++++\big\rangle &=&
  \overline{\partial}_{2}\,\partial_3~
  {\cal A}_{6,12}\big|_{\eta \to 0}
  ~=~-\frac{\langle12\rangle^3\langle13\rangle}{\<12\>\<23\>\ldots\<61\>}\,,
  \\
  \nonumber
  \big\langle\lambda^-\lambda^-\lambda^+\lambda^+++\big\rangle &=& \overline{\partial}_{1}\,\overline{\partial}_{2}\,
  \partial_3\,\partial_4~
  {\cal A}_{6,12}\big|_{\eta \to 0} ~=~
  -\frac{\langle12\rangle^3\langle34\rangle}{\<12\>\<23\>\ldots\<61\>}\,.
\end{eqnarray}
The equivalent calculations in the $\cn=4$ formalism read
\begin{eqnarray*}
  \big\langle--++++\big\rangle
  &=&
  (\partial_1^1\partial_1^2\partial_1^3\partial_1^4)
  (\partial_2^1\partial_2^2\partial_2^3\partial_2^4)
  {\cal A}_6^{\rm{MHV}} = \frac{\langle12\rangle^4}{\<12\>\<23\>\ldots\<61\>} \,,
  \\
   \big\langle-\lambda^{234}\lambda^1+++\big\rangle
   &=&
  (\partial_1^1\partial_1^2\partial_1^3\partial_1^4)
  (\partial_2^2\partial_2^3\partial_2^4) (\partial_3^1)
  {\cal A}_6^{\rm{MHV}} =
  -\frac{\langle12\rangle^3\langle13\rangle}{\<12\>\<23\>\ldots\<61\>}\,,
  \\
  \big\langle\lambda^{234}\lambda^{234}\lambda^1\lambda^1++\big\rangle
  &=&
   (\partial_1^2\partial_1^3\partial_1^4)
  (\partial_2^2\partial_2^3\partial_2^4) (\partial_3^1) (\partial_4^1)
 {{\cal A}}_6^{\rm{MHV}} =
  -\frac{\langle12\rangle^3\langle34\rangle}{\<12\>\<23\>\ldots\<61\>}\,.
\end{eqnarray*}
They agree with the $\cn=1$ results (\ref{testresult}).

An alternative form of the $\cn=1$ SYM
\gt~is
\begin{equation}\label{mhv delta tilde}
    \mathcal{A}^{\text{MHV}}_{n,ij}
    =\frac{\langle ij\rangle^3}
     {\cyc(1\ldots n)}
     \tilde{\del}\text{(ext)}\,\bh_{i}\,\bh_{j}
     \,, 
\end{equation}
where
 \be\label{delta tilde}
 \tilde{\del}(\text{ext}) =
 \tilde{\del}\Big{(}\sum\limits_{k=1}^n
|i\>\x_i\Big{)}=\frac{1}{2}\sum\limits_{k,l=1}^n\<kl\>\x_k\x_l,
~~~~~\x_k=
  \begin{cases}
\h_k & ~~~\text{if} ~~\F_k \\
-\bar{\partial}_k & ~~~\text{if} ~~\F_k^\dagger
\end{cases} \\
     \,.
\ee
This representation is homogeneous in the $\xi_i$'s, and it is easier to use in calculations.

One can formulate super-BCFW recursion relations in the $\Phi$-$\Phi^\dagger$ formalism and use it to derive N$^K$MHV superamplitudes for $\cn=1$ SYM. We have solved these relations explicitly at the NMHV level as a healthy exercise, and the result is similar to that of $\cn=4$ SYM \cite{Drummond:2008cr}. We spare the reader for details since we will shortly introduce a more convenient formalism.

\subsection{MHV vertex expansion}
\label{s:CSW}

The simple scaling argument given in \cite{Cohen:2010mi} proves that the MHV vertex expansion is valid for all tree amplitudes in $\cn=1$ SYM. In the superamplitude formalism, the MHV vertex diagrams consist of MHV superamplitudes `glued' together with propagators $1/P_I^2$ and a sum over possible states of the internal line. This sum is carried out in $\cn=4$ SYM as the simple fourth order Grassmann differentiation 
(or, equivalently, integration) $\prod_{a=1}^4 \tfrac{\partial}{\partial\eta_{Pa}}$ of the $\eta_{Pa}$'s associated with the internal line. This automatically takes care of the internal sum.
For $\cn=1$ SYM we reverse-engineer the equivalent sum as follows.

Consider a simple diagram with two MHV vertices:\\[-2mm]
\bea
    \raisebox{-9mm}{
    \includegraphics[height=1.8cm]{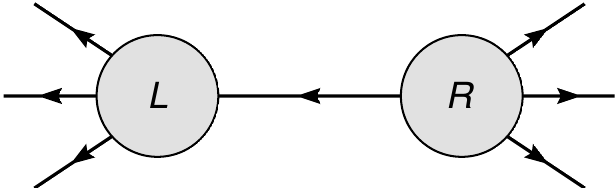}
    }
   \begin{picture}(0,0)(0,0)
   \put(-65,-1){$\Phi_i^\dagger$}
   \put(-62.5,7.5){$\Phi_j^\dagger$}
   \put(0,-1){$\Phi_k^\dagger$}
   \put(-27.5,1.7){$\Phi_{P}^\dagger$}
   \put(-39,1.7){$\Phi_{-P}$}
   \end{picture}
\eea
We assume that the external lines $i,j,k$ are in the $\Phi^\dagger$-sector and all other lines are $\Phi$'s, as appropriate for an NMHV amplitude.
 Since we label both MHV superamplitudes in terms of outgoing particles, the internal line $P_I$ propagates a $\Phi^ \dagger$-state to a $\Phi$ state (and vice versa):  if the left subamplitude has a positive helicity gluon on the internal line, it will be a negative helicity gluon on the right subamplitude. Similarly for the gluinos. There are no other possibilities in pure $\mathcal{N}=1$ SYM, 
so the rule for the internal line is
\bea
  \text{internal sum} =
   (1+ \partial_P \bar{\partial}_P )  ~
   \ca_L(\Phi_i^\dagger\dots \Phi_j^\dagger\dots\Phi_{-P}\dots)
   ~\ca_R(\dots\Phi_{P}^\dagger\dots \Phi_k^\dagger)
   \Big|_{\eta_P,\bar{\eta}_P \to 0}
\eea
The first term ``1" encodes the internal gluon state and the second term $\partial_P\bar{\partial}_P$ the internal gluino. The expression can be rewritten as $ \int d\eta_P d\bar{\eta}_P (1+ \bar{\eta}_P \eta_P ) \ca_L \ca_R $. Promoting the prefactor to an exponential we find
\bea
 \text{internal sum}=
   \int d\eta_P d\bar{\eta}_P~e^{ \bar{\eta}_P \eta_P }~
    \ca_L(\Phi_i^\dagger\dots \Phi_j^\dagger\dots\Phi_{-P}\dots) ~\ca_R(\dots\Phi_{P}^\dagger\dots \Phi_k^\dagger)
\eea
$\ca_L$ is independent of $\bar{\eta}_P$, so we can move the exponential and the $\bar{\eta}_P$-integral to act only on $\ca_R$, where it becomes the inverse Fourier transform of the state $\Phi_P^\dagger$. We note that
\bea
   \int d\bar{\eta}~e^{ \bar{\eta}\eta }~\Phi^\dagger
   ~=~ \lambda^- + \eta \, G^-
   ~\equiv~ \Psi \, ,
   \label{1stPsi}
\eea
and hence we can write
\bea
   \text{internal sum} =
   \int d\eta_P~
   \ca_L(\dots\Phi_{-P}\dots) ~\ca_R(\dots\Psi_{P} \dots) \, .
\eea
This is the simple $\cn=1$ SYM analogous of the $\cn=4$ internal line Grassmann integral.

We can convert all $\Phi^\dagger$'s in the superamplitudes to $\Psi$'s by a inverse Fourier transformation. The resulting $\Phi$-$\Psi$ formalism only depends on $\eta$'s and not $\overline\eta$'s, and this is more convenient for practical calculations than the perhaps more intuitive $\Phi, \Phi^\dagger$ formalism.

\section{Pure $\cn=0,1,2,3,4$ SYM: $\Phi$-$\Psi$ formalism}
\label{s:PhiPsi}

It was observed in \cite{Bern:2009xq}
that the unitarity cuts of pure $\cn<4$ SYM are subsets of the  $\cn=4$ cuts. In particular, when the $\cn\!=\!4$  $\eta$-integrals are converted into index diagrams  \cite{Bern:2009xq}, the  $\cn<4$ super-sum corresponds to the subset of diagrams where the $4-\cn$ index lines are grouped together. This can be understood as the embedding of the on-shell states of the $4-\cn$ theories in the maximal multiplet.  Thus one can obtain the $\cn<4$ amplitudes from the maximally SUSY ones by simply separating out the needed $\cn$ $\eta$'s. This is implemented by either integrating out, or setting to zero, the remaining $4\!-\!\cn$ $\eta$'s. This gives the  $\Phi$-$\Psi$ formalism which we now study in detail.

Let us first recall that $\Phi = G^+ + \lambda^+ \eta$ was obtained from the $\cn=4$ superfield $\Phi$ of \reef{n4Phi} by setting $\eta_{2,3,4} \to 0$, and dropping the subscript $1$. We can obtain $\Psi=\lambda^- + \eta \, G^-$ from \reef{n4Phi}  by integrating over $\eta_{2,3,4}$. (This gives the same result as \reef{1stPsi}.) The higher-$\cn$ generalizations should be clear, and we find that the on-shell states of the pure SYM theories are nicely packaged as
\begin{equation}
\begin{array}{llll}
  & \;\;\F_{\cn=1}=\Phi|_{\eta_2,\eta_3,\eta_4\rightarrow0}   ~~~&
  {\Psi}_{\cn=1}=\int d\eta_2\,d\eta_3\,d\eta_4~\Phi \,,\\[3mm]
&
\;\;\F_{\cn=2}=\Phi|_{\eta_3,\eta_4\rightarrow0}
&{\Psi}_{\cn=2}=\int d\eta_3\,d\eta_4~\Phi\,, \\[3mm]
 &
\;\;\F_{\cn=3}=\Phi|_{\eta_4\rightarrow0}
&{\Psi}_{\cn=3}=\int d\eta_4~\Phi\,.
\end{array}
\end{equation}
Explicitly, we have
\begin{equation}
\begin{array}{llll}
\mathcal{N}=1&\rightarrow&
\begin{array}{l}
 \F_{\cn=1}\,=\,G^++\eta\lambda^{+} \\[2mm]
 {\Psi}_{\cn=1}\,=\,\lambda^{-}+\eta \,G^-
\end{array}
~~~~~~~~~~{\scriptstyle (\l^+ = \l^1\,, ~~~~\l^- = \l^{234})}
\\[8mm]
  \mathcal{N}=2&\rightarrow&
\begin{array}{l}
  \F_{\cn=2}\,=\,G^++\eta_a\,\lambda^{a+}- \eta_1 \eta_2 \, S \\[2mm]
 {\Psi}_{\cn=2}
  \,=\,
  \bar{S} + \eta_a\,\lambda^{a-} - \eta_1 \eta_2\,  G^-
\end{array}
~~~~~~~~~~{\scriptstyle (\bar{S} = S^{34}\,,~~~~\l^{a-} = \l^{a34})}
\\[8mm]
 \mathcal{N}=3&\rightarrow&
\begin{array}{l}
   \F_{\cn=3}\,=\,
   G^+
   + \eta_a\, \lambda^{a}
   -\eta_a\eta_b\,S^{ab}
   -\eta_1 \eta_2 \eta_3\,\lambda^{123} \\[2mm]
  {\Psi}_{\cn=3}\,=\,
   \lambda^{4} + \eta_a\, S^{a4}
   +\frac{1}{2!} \eta_a\eta_b\,\lambda^{ab4}_c - \eta_1 \eta_2 \eta_3 \,G^-
\end{array}
\end{array}
\label{N123}
\end{equation}

The Grassmann operators associated with each state can be read-off 
from the superfields, just as we did in the $\cn=4$ case. For example, a negative helicity gluon is projected out from $\Psi_{\cn=1}$ by $\partial$, from $\Psi_{\cn=2}$ by $+\partial^1\partial^2$ and from $\Psi_{\cn=3}$ by $+\partial^1\partial^2\partial^3$.

\vspace{3mm}
\noindent {\bf Equivalence of $\cn=3$ and $\cn=4$ SYM:}\\
Let us compare the  $\cn=3$ superfields in \reef{N123} with the $\cn=4$ self-conjugate superfield (\ref{n4Phi}) with $\eta_4$ separated out:
\bea
  \nonumber
  \Omega
    &=& G^++\h_a\l^{+a}-\frac{1}{2!}\h_a\h_b S^{ab}-\frac{1}{3!}\h_a\h_b\h_c
    {\l}^{abc}\\
    &&~~~\,
     ~ +~\h_4\,\big(\l^{4} + \h_a S^{a4}
     +\frac{1}{2!}\h_a\h_b{\l}^{ab4} - \eta_1 \eta_2 \eta_3 \,
     G^-\big)\,,
\eea
with $a,b,c=1,2,3$.
We immediately recognize that
$\Omega =  \F_{\cn=3} + \eta_4 \, {\Psi}_{\cn=3}$, i.e.~the field content of the $\cn=3$ superfields \reef{N123} is equivalent to that of  $\cn=4$ SYM.
This is no surprise since $\cn=3$ SYM is equivalent to $\cn=4$ SYM.
When we apply the on-shell formalism for $\cn<4$ SYM in the following, we will occasionally compare the results of the $\cn=3$  formulation with that of $\cn=4$.

\subsection{MHV superamplitudes for $0\le \mathcal{N} \le4$}
\label{s:MHV01234}
Consider the $\mathcal{N}=1$ MHV amplitude. Choosing the $i$th and $j$th particles to be in the $\Psi$ sector, one derives the $\mathcal{N}=1$ amplitude by  integrating away 3$\eta_i$'s and 3$\eta_j$'s from the $\mathcal{N}=4$ result:
\eq
  \label{N1MHV}
  \cf_{n,ij}^{\mathcal{N}=1}
  ~=~\int d^3\eta_id^3\eta_j
  \frac{ \delta^{(8)} \big(\sum |k\> \eta_k \big)}
     {\langle12\rangle\langle 23\rangle \cdot\cdot \langle n\;1\rangle}
     \bigg|_{\eta_{k,\{2,3,4\}}\to 0}
     ~=~
  \frac{\langle ij\rangle^3~\delta^{(2)}\big(\sum |k\> \eta_k\big)}{\langle12\rangle\langle 23\rangle \cdot\cdot \langle n\,1\rangle}\,.
 \eqe
Here we have used $d^3\eta_i d^3\eta_j
=d\eta_{i2}d\eta_{i3}d\eta_{i4} d\eta_{j2}d\eta_{j3}d\eta_{j4}
= - d\eta_{i2}d\eta_{j2} d\eta_{i3}d\eta_{j3} d\eta_{i4}d\eta_{j4}$.
Each $d\eta_{ia}d\eta_{ja}$ projects out $-\<ij\>$, so all in all we get $-(-\<ij\>)^3 = \<ij\>^3$.

Similar, one finds the $\cn=0,1,2,3,4$ MHV superamplitude to be
\eq
  \label{MHVgenN}
  \cf_{n,ij}^{\mathcal{N}}
 ~=~
  (-1)^{\frac{1}{2}\cn(\cn-1)} ~\frac{\langle ij\rangle^{4-\cn}~\delta^{(2\cn)} \big(\sum |k\> \eta_k \big)}{\langle12\rangle\langle 23\rangle \cdot\cdot \langle n\,1\rangle}\,.
 \eqe
Note that $\mathcal{N}=3$ encodes $\cn=4$ processes in which the particles on lines $i$ and $j$ have been chosen to always carry $SU(4)$ index 4 while particles on all other lines  never carry  index 4. Thus the ${n \choose 2}$ different superamplitudes $\cf_{n,ij}^{\cn=3}$ encode exactly the same processes as the $\mathcal{N}=4$ superamplitude.

To obtain component amplitudes from the superamplitudes $\cf_{n,ij}^\cn$, one selects the superamplitude with superfields arranged according to the desired external states.
For example, the $\cn=1$ SYM amplitude $\langle-\lambda^-\lambda^++++\rangle$ is projected out from the MHV superamplitude $\cf^{\cn=1}_{6,12}$. The only tricky part is to keep track of the overall sign of the amplitude. To illustrate the issue, consider how to obtain the following three  $\cn=1$ amplitudes from the $\cn=4$ constructions:
\bea
  \nonumber
  \< +--++\> &=&
  (\partial_2^1 \partial_2^2 \partial_2^3 \partial_2^4)
  (\partial_3^1 \partial_3^2 \partial_3^3 \partial_3^4) \ca_5^{\cn=4}
  ~=~
  - \partial_2^1 \partial_3^1 \big[( \partial_2^2 \partial_2^3 \partial_2^4)(\partial_3^2 \partial_3^3 \partial_3^4) \ca_5^{\cn=4}\big]\\
  \label{sign1}
  &=&- \partial_2^1 \partial_3^1 \,\cf_{5,23}^{\cn=1} \,,
  \\ [2mm]
  \nonumber
   \< +  \l^- -  \l^+ +\> &=&
  (\partial_2^2 \partial_2^3 \partial_2^4)
  (\partial_3^1  \partial_3^2 \partial_3^3 \partial_3^4) \partial_4^1 \ca_5^{\cn=4}
  ~=~
  -\partial_3^1 \partial_4^1 \big[( \partial_2^2 \partial_2^3 \partial_2^4)(\partial_3^2 \partial_3^3 \partial_3^4) \ca_5^{\cn=4}\big]\\
  \label{sign2}
  &=& -  \partial_3^1 \partial_4^1 \,\cf_{5,23}^{\cn=1} \,,
  \\ [2mm]
  \nonumber
   \< +  - \l^- \l^+ +\> &=&
  (\partial_2^1 \partial_2^2 \partial_2^3 \partial_2^4)
  ( \partial_3^2 \partial_3^3 \partial_3^4) \partial_4^1 \ca_5^{\cn=4}
  ~=~
  \partial_2^1 \partial_4^1 \big[( \partial_2^2 \partial_2^3 \partial_2^4)(\partial_3^2 \partial_3^3 \partial_3^4) \ca_5^{\cn=4}\big]\\
  &=& +  \partial_2^1 \partial_4^1 \,\cf_{5,23}^{\cn=1} \,.
  \label{sign3}
\eea
Recall that the $\cn=1$ projection rules are
\bea
  \label{N1rules}
  \Phi\!:~~~~G^+ \lra 1\,,~~~~~
  \l^+ \lra \partial_i^1\,,~~~~~~~~
  \Psi\!:~~~G^- \lra \partial_i^1\,,~~~~~
  \l^- \lra 1\,.
\eea
The first two cases \reef{sign1}-\reef{sign2} require a minus sign in addition to the projection rules \reef{N1rules}. This arises from anti-commuting $\partial^{2,3,4}$'s all the way to the right. We can take this into account by the

\vspace{2mm}
\emph{Sign Rule: in the $\Phi$-$\Psi$ formalism for $\cn=1,3$ SYM one must include a minus sign everytime a Grassmann derivative moves past a $\Psi$-state}.

\vspace{2mm}
In the example \reef{sign1}, $\partial^{1}_3$ has to move past $\Psi_2$ to hit $\Psi_3$, and the Sign Rule tells us to include the overall minus sign. In the second  example, \reef{sign2},  
$\partial^{1}_4$ moves past both $\Psi_2$ and $\Psi_3$ while  $\partial^{1}_3$ has to move past $\Psi_2$; this gives an overall minus sign. In the final case \reef{sign3}, the Grassmann derivatives move past an even number of $\Psi$'s, so the Sign Rule gives "+".

\vspace{3mm}
For $\cn=2$ SYM, let us for example consider the 6-point amplitudes
$\langle\lambda^{-}_1\lambda^-_2\lambda^{+1}\lambda^{+2}++\rangle$.
and
$\langle\lambda^{-}_1\lambda^-_2+ S ++\rangle$.
These come from the MHV superamplitude $\cf_{6,12}^{\cn=2}$ in \reef{MHVgenN}.
  We apply the operators corresponding to the external states and find
\eqa
\nonumber\langle\lambda^{1-} \lambda^{2-}\lambda^{1+}\lambda^{2+}++\rangle&=&
(\partial_{1}^1)\,(\partial_{2}^2)\,(\partial_{3}^1)\,(\partial_{4}^2)\;
\cf_{6,12}^{\cn=2}
~=~\frac{\langle 12\rangle^2\langle 13\rangle\langle 24\rangle}{\langle12\rangle\langle 23\rangle \cdot\cdot \langle 61\rangle}
~
\\[2mm]
\nonumber\langle\lambda^{1-} \lambda^{2-} +S++\rangle&=&
(\partial_{1}^1)\,(\partial_{2}^2)\,(\partial_{4}^1\partial_{4}^2)\;
\cf_{6,12}^{\cn=2}
~=~\frac{\langle 12\rangle^2\langle 14\rangle\langle 24\rangle}{\langle12\rangle\langle 23\rangle \cdot\cdot \langle 61\rangle}~
\eqae
These can be seen to agree with the equivalent amplitudes obtained in the $\cn=4$ formalism.

\subsection{NMHV superamplitudes for $0\le \mathcal{N} \le4$}\label{NMHVS}

We start with the dual superconformal form of the $\mathcal{N}=4$ NMHV amplitude derived in~\cite{Drummond:2008cr}. It is expressed in terms of variables $x_i^{\alpha\dot{\alpha}}$, $|i\>^\alpha$, $\eta_i^a$, where the `region variables' $x_i^{\alpha\dot{\alpha}}$ are related to the momenta via
\eq
x_i^{\alpha\dot{\alpha}}-x_{i+1}^{\alpha\dot{\alpha}}=p_i^{\alpha\dot{\alpha}} \,.
\eqe
The $\mathcal{N}=4$  NMHV superamplitude is given as
\eqa
\label{NMHVn4}
\mathcal{N}=4:~~~~\;\; \mathcal{A}_n^\text{NMHV}=\frac{ \delta^{(8)}(\sum |k\> \eta_k)}{\langle12\rangle\langle 23\rangle \cdot\cdot \langle n\;1\rangle}\sum_{2\leq s< t\leq n-1}R_{nst}
\,,
\eqae
where
\eqa
R_{nst}&=&\frac{\langle s\;s-1\rangle\langle t\;t-1\rangle
~\prod_{a=1}^4 \Xi_{nst,a}
}
{x^{2}_{st}\langle n|x_{ns}x_{st}|t\rangle\langle n|x_{ns}x_{st}|t-1\rangle\langle n|x_{nt}x_{ts}|s\rangle\langle n|x_{nt}x_{ts}|s-1\rangle} \, ,
\eqae
 and the Grassmann odd function $\Xi_{nst,a}$ is defined as
 \eq
 \Xi_{nst,a}\equiv\sum_{i=t}^{n-1}\langle n|x_{ns}x_{st}|i\rangle \eta_{ia}+\sum_{i=s}^{n-1}\langle n|x_{nt}x_{ts}|i\rangle \eta_{ia}\,.
 \label{XiDef}
 \eqe

To derive superamplitudes for $\cn<4$ SYM, we simply perform the integrals $\int d^{4-\cn}\eta$ of $\mathcal{A}_n^\text{NMHV}$  for the three $\Psi$-states which we choose to be $i,j,n$. The details of the derivation are given in appendix \ref{app:NMHV}, here we simply state the result valid for $\cn=0,1,2,3,4$:
\eqa
&&
\label{NMHVresN}
\hspace{-10mm}
\;\mathcal{F}^\cn_{n,ijn}=
(-1)^{\frac{1}{2}\cn(\cn+1)}\,\mathcal{F}^\cn_{n,ij}\times
\\[2mm]\nonumber
&&
\Bigg[\sum_{i<s\leq j<t\leq n-1}
\Big(\langle in\rangle \langle n|x_{nt}x_{ts}|j\rangle \Big)^{4-\cn}
R^{\,\mathcal{N}}_{nst}
+\sum_{i<s<t\leq j \leq n-1} \Big(\langle ni\rangle\langle nj\rangle x^2_{st} \Big)^{4-\cn}
R^{\,\mathcal{N}}_{nst}
\\
\nonumber
&& +\sum_{2\leq s\leq i<j<t\leq n-1} \Big(\langle ij\rangle\langle n|x_{nt}x_{ts}|n\rangle \Big)^{4-\cn}R^{\,\mathcal{N}}_{nst}
+\sum_{2\leq s\leq i<t\leq j} \Big(\langle jn\rangle\langle n|x_{ns}x_{st}|i\rangle \Big)^{4-\cn}R^{\,\mathcal{N}}_{nst}\Bigg],
\eqae
 where
 \eq
 R^{\mathcal{N}}_{nst}=\frac{\langle s\;s-1\rangle\langle t\;t-1\rangle
 \prod_{a=1}^\mathcal{\cn} \Xi_{nst,a}}{x^{2}_{st}\langle n|x_{ns}x_{st}|t\rangle\langle n|x_{ns}x_{st}|t-1\rangle\langle n|x_{nt}x_{ts}|s\rangle\langle n|x_{nt}x_{ts}|s-1\rangle}\,.
 \label{RN}
 \eqe
For $\cn=0$ SYM, the product in \reef{RN} is set to 1; this result was presented recently in \cite{Dixon:2010ik}.


\subsection{On the range of $\cn$ in SYM}
We have derived MHV and NMHV superamplitudes $\cf^\cn_{n,ij}$ and
$\cf^\cn_{n,ijk}$ in which the number of supersymmetries $\cn$ appeared as a parameter. We know the interpretation of these superamplitudes for $\cn=0,1,2,3,4$, but what if $\cn$ takes other (integer) values? Clearly, $\delta^{(2\cn)}$ makes sense only for $\cn\ge 0$. For $\cn>4$, $\cf^\cn_{n,ij\dots}$ is not a physical object. To see this, let us just consider $\cn=5$.

For $\cn=5$, the tree level MHV superamplitude $\cf^{\cn=5}_{n,ij}$ in \reef{MHVgenN} includes an amplitude
\bea
  \label{badA}
  \< A \dots \> = \frac{\<ij\>^{-1} \<ab\>^5}{\<12\> \dots \<n1\>}
\eea
where $a,b \ne i$. Under a little group scaling of line $i$, the amplitude \reef{badA} scales as $t^{-3}$, so this immediately tells us that line $i$ is a particle with helicity $3/2$.
This should already raise suspicion since it is also easy to see that there are no spin 2 particles possible within the same superamplitude. Now if lines $i$ and $j$ are non-adjacent, \reef{badA} has a pole in the $ij$-momentum channel. This is unphysical because the amplitude is color-ordered. If the $ij$ are adjacent, then
\reef{badA} $\<ij\>$ already appears in the cyclic product of angle brackets, and hence there is a double-pole in the $ij$-momentum channel; this is also unphysical. We conclude that $\cf^{\cn=5}_{n,ij}$ (or $\cf^{\cn>4}_{n,ij}$) does not encode  sensible tree amplitudes of a local non-gravitational field theory.

In supergravity,  $\cn$ can take a larger range of values; we will discuss briefly the $0\le \cn\le 8$ supergravity superamplitudes in section \ref{s:SG}.

\section{Super-BCFW}
\label{s:bcfw}

The super-BCFW shift, introduced for the maximally supersymmetric theories $\cn=4$ SYM (and $\cn=8$ supergravity) in \cite{ArkaniHamed:2008gz,Brandhuber:2008pf} is\footnote{There is also a super-shift relevant for the MHV vertex expansion, see \cite{Kiermaier:2009yu}.} 
\bea
\nonumber
 \text{$\cn=4$ SYM:}~~~&&
   |\hat{I}] = |I] + z\, |J] \, ,\qquad
   |\hat{J}\> = |J\> - z\, |I\>\, ,\\
   &&
   \hat{\eta}_{Ia}= \eta_{Ia} + z \eta_{Ja} ~~\text{for}~~a=1,2,3,4 \, .
 \label{n4bcfw}
\eea
Under this shift, the tree level $\cn=4$ superamplitudes behave as
\bea
  \label{n4largez}
    \ca_n^{\cn=4} \sim 1/z  ~~~\text{as}~~z\to\infty
\eea
when lines $I$ and $J$ are adjacent, and as $1/z^2$ when they are non-adjacent.

The large-$z$ falloff implies a set of valid recursion relations for superamplitudes. These recursion relations were solved in \cite{Drummond:2008cr} to yield dual superconformal invariant expressions for any tree-level N$^K$MHV superamplitudes of $\cn=4$ SYM. This includes the $\cn=4$ NMHV superamplitude expressions used in section \ref{NMHVS}.

In this section, we generalize the super-BCFW shift to $\cn<4$ SYM and discuss its validity. When valid, the super-BCFW recursion relations can be solved just as in $\cn=4$ SYM; however, as we have shown how to truncate the $\cn=4$ SYM tree results to  $\cn<4$ SYM, there is no need to pursue this direction. The important outcome of this section therefore is to characterize when the super-BCFW shifts have large-$z$ falloff and when that fails. This will have influence of the 1-loop structure of the amplitudes, as we discuss in section \ref{s:1loop}.

We work in the $\Phi$-$\Psi$ formalism.
To be specific, we specialize to $\cn=1$ SYM, but our discussion and results generalize directly to $\cn=2$ and $\cn=3$. Consider a $[I,J\>$ super-BCFW shift
\bea
 \label{n1bcfw}
   |\hat{I}] = |I] + z\, |J] \, ,\qquad
   |\hat{J}\> = |J\> - z\, |I\>\, ,\qquad
   \hat{\eta}_I = \eta_I + z \eta_J \, .
\eea
All other spinors and $\eta$'s are unshifted.
By construction, the shift \reef{n1bcfw} leaves the Grassmann $\delta$-function $\delta^{(2)}(\tilde{Q})$ 
invariant. It only takes a moment of inspection to realize that the MHV amplitude \reef{N1MHV} behaves as
\bea
\label{4shifts}
  \begin{array}{ccccccc}
  [\Psi,\Psi\>  &~~&
  [\Phi,\Phi\> &~~&
  [\Psi,\Phi\> &~~&
  [\Phi,\Psi\>\\[1mm]
  1/z &&1/z &&1/z && z^2
  \end{array}
\eea
for large $z$ under the adjacent super-BCFW shift \reef{n1bcfw}.
 We have indicated to which sectors the two shifted lines belong. For shifts of non-adjacent lines, the falloff is a factor of $1/z$ better than in \reef{4shifts}.\footnote{Note that the behavior mimics that of gluon amplitudes under regular BCFW $[\pm,\pm\>$ and $[\mp,\pm\>$ shifts, with only a small improvement $z^3 \to z^2$ thanks to the $\cn$=1 supershift.}

The large-$z$ behavior \reef{4shifts} is valid also for N$^K$MHV tree superamplitudes. To show this, consider a general N$^K$MHV superamplitude of $\cn=1$ SYM; it has $(K\!+\!2)$ $\Psi$-lines and the
$(n\!-\!K\!-\!2)$ $\Phi$-lines. The $\cn=1$ superamplitude $\cf_n$ is obtained from that
 of $\cn=4$ SYM as
\bea
  \cf_{n} &=& \bigg[
  \int \Big( \prod_{a=2}^4 \prod_{x \in \Psi}\, d\eta_{xa} \Big)~\ca_n^{\cn=4}
  \bigg]_{\eta_{y2},\eta_{y3},\eta_{y4} \to 0~\text{for}\,\,y \in \Phi}
  \,.
\eea
The truncation rule ${\eta_{y2},\eta_{y3},\eta_{y4} \to 0~\text{for}\,\,y \in \Phi}$ can be converted an Grassmann integration by integrating over all $\eta_{2,3,4}$'s with a `measure' containing the product of all $\eta_{ya}$'s for $a=2,3,4$:
\bea
  \cf_{n} &=&
  \int \Big( \prod_{a=2}^4 \prod_{i=1}^n\, d\eta_{ia}  \prod_{y\in \Phi} \,\eta_{ya} \Big)~\ca_n^{\cn=4} \, .
\eea
When we apply the $\cn\!=\!1$ supershift \reef{n1bcfw}, it only acts on $\eta_{I1}$ in $\ca_n^{\cn=4}$, i.e.~the shifted superamplitude $\ca_n^{\cn=4}$ depends on $\hat{\eta}_{I1} = \eta_{I1}+z\eta_{J1}$ and on $\eta_{Ia}$ for $a=2,3,4$. To use the result \reef{n4largez} for the large-$z$ falloff of $\ca_n^{\cn=4}$, we need all four $\eta_{Ia}$ to be shifted. To accomplish this,  we redefine for $a=2,3,4$ the integration variables as
\bea
  \eta_{Ia} \to
  \tilde{\eta}_{Ia} - z\tilde{\eta}_{Ja}\,,
  ~~~~\text{and}~~~~
  \eta_{ia} \to \tilde{\eta}_{ia}~~~  \text{for all}~~~ i\ne I\,.
\eea
  The Jacobian is 1, so we can  write the shifted $\cn=1$ superamplitude
\bea
  \hat{\cf}_{n}(z)
  &=&
  \int \Big( \prod_{a=2}^4 \prod_{i=1}^n\, d\tilde{\eta}_{ia} \, \hat{m}_a \Big)~\hat{\ca}_n^{\cn=4}(z) \, ,
  ~~~~\text{with}~~~~
  \hat{m}_a = \prod_{y\in \Phi} \,\eta_{ya}(\tilde{\eta}_{ia}) \, .
\eea
 Note that $\hat{}$ on $\cf$ indicates the $\cn=1$ supershift \reef{n1bcfw} while
$\hat{}$ on $\hat{\ca}_n^{\cn=4}$ refers to a full $\cn=4$ supershift \reef{n4bcfw}, thanks to the coordinate transformation in the integral. We already know that for large $z$, $\hat{\ca}_n^{\cn=4}(z)$ goes as $1/z$ (or better), so the only way the large-$z$ behavior of
$\hat{\cf}_{n}(z)$ can differ is if the `measure'-factor $\hat{m}_a$ shifts. Let us go through the four different shifts and track the large-$z$ behavior:
\begin{itemize}
\item $[\Psi,\Psi\>$ and $[\Psi,\Phi\>$: when $I \notin \Phi$, all
$\,\eta_{ya}(\tilde{\eta}_{ia}) = \tilde{\eta}_{ya}$; they are $z$-independent, so $\hat{m}_a \sim O(1)$ for large $z$.
\item $[\Phi,\Phi\>$: when $I,J \in \Phi$, the factor  $\hat{m}_a$ contains
both $\eta_{Ia}$ and $\eta_{Ja}$. Their product is $(\tilde{\eta}_{Ia} -z \tilde{\eta}_{Ja})\,\tilde{\eta}_{Ja} = \tilde{\eta}_{Ia} \,\tilde{\eta}_{Ja}$, so $\hat{m}_a\sim O(1)$ for large $z$.
\item $[\Phi,\Psi\>$: in this case  $I\in \Phi$, and hence  $\hat{m}_a$ contains a factor of $\eta_{Ia} \to (\tilde{\eta}_{Ia} -z \tilde{\eta}_{Ja})$. But there is no factor of  $\tilde{\eta}_{Ja}$ in  $\hat{m}_a$ because $J\in \Psi$, so we conclude that $\hat{m}_a \sim z$ for large $z$. The three factors $a=2,3,4$ of $\hat{m}_a$ thus give a large $z$ behavior of $z^3$.
\end{itemize}

Together with the result \reef{n4largez} that $\hat{\ca}_n^{\cn=4}(z) \sim 1/z$ for large $z$ for a shift of adjacent lines, we conclude that \reef{4shifts} indeed holds for all $\cn=1$ superamplitudes.
The generalization of this result to $\cn=2,3$ follows from a similar argument, but with $4-\cn$ factors in the Grassmann integration. The general result can be summarized as
\bea
\label{4shiftsN}
  \begin{array}{ccccccc}
  [\Psi,\Psi\>  &~~&
  [\Phi,\Phi\> &~~&
  [\Psi,\Phi\> &~~&
  [\Phi,\Psi\>\\[1mm]
  1/z &&1/z &&1/z && z^{3-\cn}
  \end{array}
\eea
This is valid for all pure  SYM N$^K$MHV superamplitudes at the tree level with $\cn=0,1,2,3,4$.

\section{Bubble contributions to 1-loop amplitudes}
\label{s:1loop}

The relationship between amplitudes in $\cn < 4$ SYM and $\cn = 4$ SYM is not as straightforward at loop-level as it is at tree level. We remarked earlier that $\cn < 4$ super-sum results can be obtained from $\cn=4$ super-sums \cite{Bern:2009xq}, but a non-trivial task is then to keep track of the relative signs of each contribution. It is more direct to use the $\cn<4$ tree superamplitudes to construct the loops. We illustrate this here by evaluating  explicitly the bubble contributions to 1-loop amplitudes in pure $\cn=1,2$ SYM.

In four dimensions, the 1-loop amplitudes can be expanded as \cite{GenUni}
\begin{eqnarray}\label{loop1}
  A^{\rm{1-loop}} &=&
  \sum\limits_i C_\text{box}^i I_\text{box}^i
  +\sum\limits_i C_\text{triangle}^i I_\text{triangle}^i
  +\sum\limits_i C_\text{bubble}^i I_\text{bubble}^i
  + \text{rationals} \,.~~~~~
\end{eqnarray}
The amplitudes of SYM theories are cut constructible at 1-loop \cite{Bern:1994zx}: there are no rational terms and the four dimensional cuts determine the full amplitude. 
The coefficients $C_p$'s are rational functions of kinematical invariants. A box coefficient $C_\text{box}$ is the product
of four on-shell tree amplitudes with the intermediate state sum carried out suitably. Triangle coefficients $C_\text{triangle}$
and bubble coefficients $C_\text{bubble}$ can be determined as
the ``pole at infinity" of the products of three and two on-shell amplitudes, respectively \cite{Forde:2007mi,ArkaniHamed:2008gz}.

The integrals $I_p^i$ in \reef{loop1} are scalar integrals of box, triangle and bubble  scalar  diagrams. Among these, only the bubble integrals have UV divergences, and hence the bubble coefficients carry information about the 1-loop beta-function \cite{Forde:2007mi,ArkaniHamed:2008gz,Lance}.  
We discuss this in section \ref{s:beta}.  Accordingly, bubbles (as well as triangles and rationals) are absent in $\cn=4$ SYM, but they yield non-vanishing contributions to the 1-loop amplitudes in $\cn=0,1,2$ SYM. Our purpose here is to clarify the structure of these bubble contributions and compute them explicitly using the tools we have developed in the previous sections.

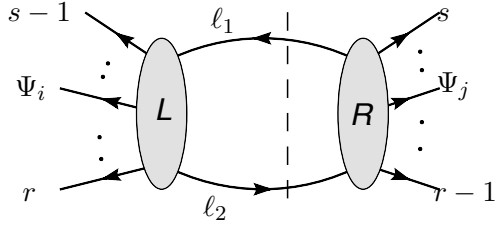
\begin{figure}[t]
\begin{center}
  \includegraphics[width=2in]{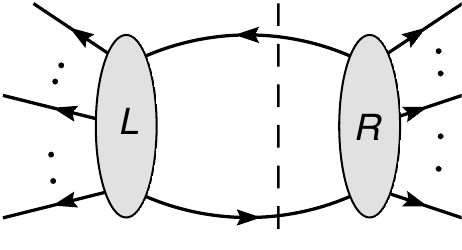}
\end{center}
\begin{picture}(0,0)(0,0)
 \put(104,25.5){$\Psi_j$}
 \put(48,25){$\Psi_i$}
 \put(74,34){$\ell_1$}
 \put(73,9){$\ell_2$}
 \put(104,35){$s$}
 \put(103.6,11){$r-1$}
 \put(47,35){$s-1$}
 \put(49,11){$r$}
\end{picture}
\vspace{-6mm}
  \caption{\small Example of 1-loop double cut.}\label{eg2}
\end{figure}

\subsection{Which bubble coefficients contribute?}
\label{s:whichbub}

Consider the 2-line cut of a 1-loop amplitude in figure \ref{eg2}.
It was shown in \cite{ArkaniHamed:2008gz} that the corresponding bubble coefficient can be calculated as\footnote{We have fixed the normalization in \reef{bubble1}  by requiring that the bubble coefficient is 1 when evaluated for a 4-point 1-loop amplitudes in color-ordered $\phi^4$-theory; see footnote \ref{footie:phi4}.}
\begin{eqnarray}\label{bubble1}
  C^\text{(L,R)}_\text{bubble} &=&
  \frac{1}{(2\pi i)^2}
  \int d \text{LIPS}[\ell_1,\ell_2]
  \int_\mathcal{C}\frac{dz}{z}\sum_\text{state sum}
  \hat{A}_L^\text{tree}\big( |\hat{\ell}_1\>,|\hat{\ell}_2]\big)\,
  \hat{A}_R^\text{tree}\big( |\hat{\ell}_1\>,|\hat{\ell}_2]\big) \,,~~~~
\end{eqnarray}
where $d$LIPS$=d^4\ell_1d^4\ell_2 \,\delta^{(+)}(\ell_1^2)\,\delta^{(+)}(\ell_2^2)\,\delta^{4}(\ell_1-\ell_2-P)$ and $P$ is the momentum going out of the left subamplitude. The contour is around the pole at infinity and the $z$-dependence in the two on-shell tree subamplitudes is exactly that of a BCFW $[\ell_2,\ell_1\>$-shift. The $z$-integral picks out the $O(1)$-term
of the large-$z$ expansion of the shifted product $\hat{A}_L \hat{A}_R$. In the on-shell superspace formulation, the amplitudes are promoted to superamplitudes and the state sum becomes a Grassmann integral:
\begin{eqnarray}\label{bubble2}
  C^\text{(L,R)}_\text{bubble} &=&
  -\frac{1}{2\pi i}
   \int d \text{LIPS}[l_1,l_2]~
    \Big[ \widehat{\cs}^\text{\,(L,R)}_{n,ij}\Big]_{O(1)\text{ as }z\to\infty} \,,
\end{eqnarray}
where  $\widehat{\cs}^\text{(L,R)}_{n,ij}$ denotes the $[\ell_2,\ell_1\>$-shift of the state sum 
\bea
  \cs^\text{(L,R)}_{n,ij}~\equiv~
  \sum_\text{state sum}  \ca_L\,\ca_R
  ~=~ \int d^\cn\h_{\ell_1} d^\cn\h_{\ell_2}~\ca_L^\text{tree}\,\ca_R^\text{tree} \,.
\eea

Changing integration variables in this integral converts the ordinary BCFW-shift acting on the amplitudes to a super-BCFW shift.\footnote{The Jacobian is 1 for this change of Grassmann integration variables.} Thus the large-$z$ behavior of superamplitudes under super-BCFW shifts have direct implications for the bubbles --- and hence potential UV divergences --- of 1-loop amplitudes. For example, the large-$z$ falloff of all tree superamplitudes of $\cn=4$ SYM and $\cn=8$ supergravity was used in \cite{ArkaniHamed:2008gz} to show that there are no bubble contributions in these theories (see also \cite{BjerrumBohr:2008ji}). Here we will use our results for the large-$z$ behavior of  super-BCFW shifts to establish which bubble coefficients vanish and which ones contribute in $\cn=1,2$ SYM.

\begin{figure}[t]
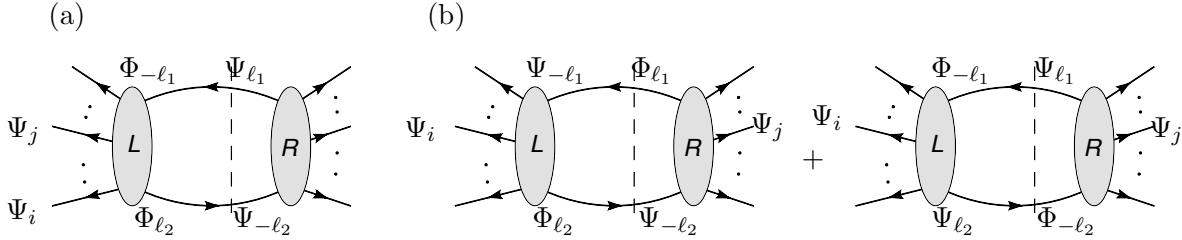

\begin{center}
 \mbox{
 \raisebox{2.7cm}{\hspace{-1mm}(a)}\hspace{-6mm}
  \includegraphics[width=4cm]{1loopCutsfig1.pdf}  ~
  \hspace{5mm}
 \raisebox{2.7cm}{(b)}\hspace{-3mm}
  \includegraphics[width=4cm]{1loopCutsfig1.pdf}  ~~~
    \raisebox{8mm}{+}~~
  \includegraphics[width=4cm]{1loopCutsfig1.pdf}}
\end{center}
\begin{picture}(0,0)(0,0)
 \put(0,10){$\Psi_i$}
 \put(0,21){$\Psi_j$}
 \put(15,29){$\Phi_{-\ell_1}$}
 \put(17,9){$\Phi_{\ell_2}$}
 \put(29,29){$\Psi_{\ell_1}$}
 \put(30,9){$\Psi_{-\ell_2}$}
 \put(99,21.5){$\Psi_j$}
 \put(53,21){$\Psi_i$}
 \put(69,29){$\Psi_{-\ell_1}$}
 \put(70,9){$\Phi_{\ell_2}$}
 \put(83,29){$\Phi_{\ell_1}$}
 \put(84,9){$\Psi_{-\ell_2}$}
 \put(107,22.5){$\Psi_i$}
 \put(152,21){$\Psi_j$}
 \put(123,29){$\Phi_{-\ell_1}$}
 \put(123,9){$\Psi_{\ell_2}$}
 \put(136.5,29){$\Psi_{\ell_1}$}
 \put(137,9){$\Phi_{-\ell_2}$}
\end{picture}
\vspace{-6mm}
  \caption{\small Two different ``bubble cuts'' of a 1-loop MHV superamplitude.}\label{fig:1loopcut}
\end{figure}

We  use the $\Phi$-$\Psi$ 
formulation of the on-shell superspace for $\cn<4$ SYM. For MHV 1-loop amplitudes $\cf_{n,ij}^\text{1-loop}$, there are two different types of bubbles, depending on whether the two external $\Psi$-states 
$i$ and $j$ belong to the same subamplitudes or not; the two cases are shown in figure \ref{fig:1loopcut}. In cuts of type (a), the super-BCFW shifts acting on the subamplitudes will be of type $[\Phi,\Phi\>$ or $[\Psi,\Psi\>$ under which we have shown in \reef{4shiftsN} 
that any $0\le\cn\le4$ superamplitude falls off as $1/z$ for large $z$. So
\bea
   \widehat{\cs}^\text{\,[Cut (a)]}_{n,ij}
   ~\sim~ \frac{1}{z} \times \frac{1}{z}
   ~\sim~ \frac{1}{z^2}~~~\text{as}~~~z \to \infty \,,
   \label{ALARz}
\eea
and hence the corresponding bubble coefficients vanish.

On the other hand, the cuts of type $(b)$ in figure \ref{fig:1loopcut} always involve a shift that acts as $[\Psi,\Phi\>$ on one subamplitude and as $[\Phi,\Psi\>$ on the other. When the internal lines are adjacent,\footnote{In non-planar amplitudes, one or more external legs can enter between the internal lines. Then the falloff \reef{ALARz} can be improved to $z^{1-\cn}$ or $z^{-\cn}$, indicating a better UV behavior of $1/N$-subleading contributions to 1-loop amplitudes in SYM.} the result \reef{4shiftsN} gives
\bea
  \label{scutN}
      \widehat{\cs}^\text{\,[Cut (b)]}_{n,ij}
   ~\sim~ \frac{1}{z} \times z^{3-\cn}
   ~\sim~ z^{2-\cn}
   ~~~\text{as}~~~z \to \infty \,.
\eea
We note  immediately  that there are no bubble contributions for $\cn=3,4$; of course this is what we expected.  The large $z$-behavior indicates that there can be non-vanishing $O(1)$-terms and hence bubble contributions for $\cn=0,1,2$. 
We now verify this by explicitly carrying out the intermediate state sum and then check the BCFW-shifts.

\subsection{Intermediate state sums}
\label{s:sums}

Let us start with $\cn=1$ SYM. In all three diagrams of figure \ref{fig:1loopcut}, the product of superamplitudes involves $\delta^{(2)}(L)\delta^{(2)}(R)$, which must be integrated over the internal Grassmann variables.
We have
\begin{equation}
\delta^{(2)}(L)=  \del^2\big{(}\sum\limits_{\text{ext $L$}}|\l\>\h_\l+|\ell_2\>\h_{\ell_{2}}-|\ell_1\>\h_{\ell_1}\big{)}\,,~~~~~
\delta^{(2)}(R)= \del^2\big{(}\sum\limits_{\text{ext $R$}}|\l\>\h_\l-|\ell_2\>\h_{\ell_{2}}+|\ell_1\>\h_{\ell_1}\big{)}\,,~~
\end{equation}
so
\begin{eqnarray}
\nonumber
 \int d\h_{\ell_1}\,d\h_{\ell_2}  ~\delta^{(2)}(L)\,\delta^{(2)}(R)
 &=&
  \int d\h_{\ell_1} \,d\h_{\ell_2} ~\delta^{(2)}(L+R)\,\delta^{(2)}(R)
\\
 \nonumber
  &=& \del^{(2)}(\text{ext})\int d\h_{\ell_1}\, d\h_{\ell_2}
  \Big( -\<\ell_1\ell_2\> \eta_{\ell_1} \eta_{\ell_2} +\dots \Big)
\\
    &=&     \del^{(2)}(\text{ext})\,\<\ell_1\ell_2\>
    \,.
    \label{intres}
\end{eqnarray}
The cases  (a) and (b) in figure \ref{fig:1loopcut} yield different results due to the different prefactors of the intermediate state sums. Case (a) gives\footnote{The overall signs, and the relative sign in case (b), are justified in appendix \ref{app:B} using the proper trunction of the $\cn=4$ state sum. The result can also be  verified by direct calculation.}
\bea
 \nonumber
  \cn=1\!:~~~~
  \cs^\text{\,[Cut (a)]}_{n,ij} &=&
  \int d\h_{\ell_1} d\h_{\ell_2}~
  \ca_L(\Psi_i \dots \Psi_j \dots \Phi_{-\ell_1} \Phi_{\ell_2})
  ~\ca_R(\dots  \Psi_{-\ell_2} \Psi_{\ell_1})\\
 \nonumber
  &=&
  \frac{\<ij\>^3\<-\ell_2,\ell_1\>^3}{\cyc(L)\cyc(R)}
   \int d\h_{\ell_1}\,d\h_{\ell_2}  ~\delta^{(2)}(L)\,\delta^{(2)}(R) \\[2mm]
  &=&
   \frac{\<ij\>^3\<\ell_1 \ell_2\>^4}{\cyc(L)\cyc(R)}\, \del^{(2)}(\text{ext})\,,
   \label{casea}
\eea
and cut (b) gives
\bea
 \nonumber
  \cn=1\!:~~~~
    \cs^\text{\,[Cut (b)]}_{n,ij} &=&
  \int d\h_{\ell_1} d\h_{\ell_2}~
  \ca_L(\Psi_i \dots  \dots \Psi_{-\ell_1} \Phi_{\ell_2})
  ~\ca_R( \dots \Psi_j \dots   \Psi_{-\ell_2} \Phi_{\ell_1})\\
 \nonumber&&+
  \int d\h_{\ell_1} d\h_{\ell_2}~
  \ca_L(\Psi_i \dots  \dots \Phi_{-\ell_1} \Psi_{\ell_2})
  ~\ca_R(\dots  \Phi_{-\ell_2} \Psi_{\ell_1}  \dots \Psi_j )\\[2mm]
  &=&
    \frac{\big(\<i \, \ell_1\>^3\<j\,\ell_2\>^3-\<i\,\ell_2\>^3\<j\,\ell_1\>^3\big)
    \<\ell_1 \ell_2\>}{\cyc(L)\cyc(R)} \, \del^{(2)}(\text{ext})\,.
    \label{c1c2}
\eea

To test the results, let us assume that all external particles are gluons, with
$i$ and $j$ the only ones with negative helicity. In {\bf (a)}, only a gluon can run in the loop, and it contributes $\<ij\>^4 \<\ell_1 \ell_2\>^4$. This matches the result \reef{casea} after the external gluons are projected out\footnote{The signs in the $\cn=1,3$ projection rules were discussed in section \ref{s:MHV01234}. }  with $-\partial_i \partial_j$, giving a factor $\<ij\>$. In the first diagram 
for case {\bf (b)}, a gluon running in the loop gives $\<\ell_1 i\>^4 \<\ell_2 j\>^4$ while a gluino gives $(-1)\<i \ell_1\>^3\<i \ell_2\> \<j \ell_2\>^3\< j \ell_1\>$ (the $(-1)$ is from the fermion loop). The sum of the two contributions Schouten to $\<i \ell_1\>^3 \<j \ell_2 \>^3\<\ell_1 \ell_2\>\<i j\>$ which matches the first term in \reef{c1c2} after extracting $\<i j\>$ from $\del^{(2)}(\text{ext})$. A similar test of the second diagram of case (b) verifies the 2nd term in \reef{c1c2}, including signs.

If we carry out the same state sums in $\cn=2$ SYM,
the $\eta$-integrals produce $-\<\ell_1\ell_2\>^{2}\del^{(4)}(\text{ext}) $. The result for the $(b)$-cut is then
\bea
   \cn=2\!:~~~~
  \cs^\text{\,[Cut (b)]}_{n,ij}
  &=&
    -\frac{\big(\<i \, \ell_1\>^2\<j\,\ell_2\>^2+\<i\,\ell_2\>^2\<j\,\ell_1\>^2\big)
    \<\ell_1 \ell_2\>^2}{\cyc(L)\cyc(R)} \, \del^{(4)}(\text{ext})\,.
\eea
Let us finally check the $\cn=3$ version:
\bea
 \nonumber
  \cn=3\!:~~~~
  \cs^\text{\,[Cut (b)]}_{n,ij}
  &\!=\!&
    -\frac{\big(\<i \, \ell_1\>\<j\,\ell_2\>-\<i\,\ell_2\>\<j\,\ell_1\>\big)
    \<\ell_1 \ell_2\>^3}{\cyc(L)\cyc(R)} \, \del^{(6)}(\text{ext})
    =
   \frac{\<i \,j\>  \<\ell_1 \ell_2\>^4}{\cyc(L)\cyc(R)} \, \del^{(6)}(\text{ext})\,.
   \\
    \label{cutbN3}
\eea
This is equivalent to the $\cn=4$ SYM cut assuming that the external states $i$ and $j$ both carry $SU(4)$ index 4; this assumption allows one to pull out $\<ij\>$ from $\del^{(8)}(\text{ext})$ leaving $\del^{(6)}(\text{ext})$.

We can summarize the result for cut $(b)$ into one formula for $\cn$-fold SYM
\bea
  \label{cutbN}
    \cs^\text{\,[Cut (b)]}_{n,ij}
   ~=~
    \frac{\Big[ \big(\<i \, \ell_1\>\<j\,\ell_2\>\big)^{4-\cn}
    +\big(\<\ell_2\,i\>\<j\,\ell_1\>\big)^{4-\cn}\Big]
    \<\ell_1 \ell_2\>^{\cn}}{\cyc(L)\cyc(R)} ~\del^{(2\cn)}(\text{ext})\, .
    ~~~~~~
\eea

Let us now assume, as in figure \ref{eg2}, that the lines $\ell_1$ and $\ell_2$ are adjacent. We label the four external lines adjacent to the internal lines by $r-1,r,s-1,s$. The state sum can then be written
\bea\label{intg}
 \cs^\text{\,[Cut (b)]}_{n,ij}
    =
     -
     \ca_{n,ij}^\text{tree} ~
     \frac{\<s-1,s\>\<r-1,r\>\,\Big[ \big(\<i \, \ell_1\>\<j\,\ell_2\>\big)^{4-\cn}
    +\big(\<\ell_2\,i\>\<j\,\ell_1\>\big)^{4-\cn}\Big]
    }{
    \raisebox{-1pt}{$\<\ell_1 \ell_2\>^{2-\cn} \,\<ij\>^{4-\cn}\, \<r \,\ell_2\>
    \<s-1 \,\ell_1\>\<s\, \ell_1\>\<r-1\,\ell_2\>$}} \,.
\eea
This formula is valid for $\cn=0,1,2,3,4$.

Let us now consider the large-$z$ behavior under a BCFW $[\ell_2,\ell_1\>$-shift. We refer to \reef{cutbN} or \reef{intg}, and note that exactly two of the angle brackets in the denominator shift. The numerators are unshifted for  $\cn=3,4$ (see \reef{cutbN3}), so the large-$z$ behavior is $1/z^2$. Note that for $\cn=3$ this is one power better than indicated by the super-shift argument \reef{scutN}. This is due to the cancellation between the contributions of the two diagrams of case (b) in figure \ref{fig:1loopcut}; such a cancellation had to take place because the $\cn=3$ and $\cn=4$ formulations are equivalent.

Applying the $[\ell_2,\ell_1\>$-shift to \reef{intg} for $\cn=1$,
one finds that the leading $O(z)$ terms predicted by \reef{scutN}
cancel between the two numerator terms; this is a
cancellation between the two diagrams in figure \ref{fig:1loopcut}.
After use of the Schouten identity, the result for the $O(1)$ terms
of $\cn=1$  and $\cn=2$ can be brought to the same form, namely
\bea
  \nonumber
  &&
  \hspace{-6mm} \cn=1,2:~~~~\\
  &&\widehat{\cs}^\text{\,[Cut (b)]}_{n,ij}
  \Big|_{O(1)} =
  -(4-\cn)\,\ca_{n,ij}^\text{tree}~
 \frac{\<s-1,s\>\<r-1,r\>}{\<ij\>^2}
  \frac{\<i \ell_2\>^2 \<j \ell_2\>^2}
  {\<r-1\, \ell_2\>\<r\, \ell_2\>\<s-1\, \ell_2\>\<s\, \ell_2\>} \, .~~~~~~~~~~
 \label{1lp}
\eea
For $\cn=0$, the $O(1)$-term takes a more complicated
form which can be found in \cite{ArkaniHamed:2008gz}. Our next task is 
to  evaluate the $d$LIPS integral \reef{bubble2} of \reef{1lp} in order to find explicit results for the bubble coefficients.

\subsection{Evaluation of the bubble cuts in $\cn=1,2$ SYM}
\label{4pt}

Let us now turn to the evaluation of the non-vanishing bubble coefficients in pure $\cn=1,2$ SYM. Following earlier work (see for example \cite{CSW,Britto:2005ha,ArkaniHamed:2008gz,Britto:2010xq}) the $d$LIPS  integral of the bubble coefficient \reef{bubble2} can be rewritten as
\bea  \label{Cbub2}
  \int d\text{LIPS}[\ell_1, \ell_2] \,(\bullet) = P^2 \int_{\tilde\lambda = \bar{\lambda}} \frac{\<\lambda, d\lambda \> [\tilde\lambda, d\tilde\lambda ]}{ \<\lambda | P | \tilde\lambda]^2} ~  \,(\bullet)  \, .
\eea
To obtain this form, $\ell_1$ has been eliminated via momentum conservation, $\ell_1 = P + \ell_2$, where $P$ is the sum of external momenta going out of $A_L$,  and $\ell_2 \propto \lambda \tilde\lambda$ \cite{ArkaniHamed:2008gz,Britto:2005ha}.

We have already established that only (b)-cuts give non-vanishing bubble coefficients.
It is clear from \reef{1lp} that the $\cn=1,2$ integrand $(\bullet)$ of \reef{Cbub2} is a rational function $g = g(\lambda)$ of angle brackets $\< x \lambda \>$ with $x$ being various external lines.\footnote{This is different from the $\cn=0$ case  \cite{ArkaniHamed:2008gz} where the state sum does not cancel a denominator factor of $\<\ell_1\ell_2\>^2$; thus in $\cn=0$, the integrand has an extra factor of $\<\l|P|\tilde\l]^{-2}$. Consequently, we follow \cite{Britto:2005ha} instead of \cite{ArkaniHamed:2008gz} when we evaluate \reef{Cbub2}.}
Following  \cite{CSW,Britto:2005ha}, we now write (in our conventions)
\bea
\int\!\!
 \frac{[\tilde\lambda, d\tilde\lambda ]}{ \<\lambda | P | \tilde\lambda]^2}
  g(\lambda)
  ~=~
  \frac{2\pi [\tilde{\lambda},\eta]}{\<\lambda|P|\tilde{\lambda}]}
  \Big\{
    - \bar{\delta}\big([\h|P|\l\>\big)\, g(\lambda)
    + \frac{1}{\<\lambda|P|\eta]}
    \sum_{k} \bar{\delta}\big(\<k,\lambda \>\big)\,
    g(\lambda)\, \<k,\lambda \>
  \Big\}\,,~~
\eea 
where we have dropped a total derivative term. 
$\eta$ is an
arbitrary reference spinor, the sum is over the simple poles
$k$ of $g$, and we have used \bea
   {d\tilde{\lambda}^\a}\frac{\partial}{\partial\tilde{\lambda}^\a} \frac{1}{\<\lambda,x\>} = 2\pi \bar{\delta}\big(\<\lambda,x\>\big)
   \, ,
   ~~~~~\text{with}~~~~~
   \int \< \lambda,d\lambda\>\,\bar{\delta}\big(\<\lambda,x\>\big)\, f(\lambda) = -i f(x).
\eea
Carrying out the $\<\lambda,d\lambda\>$-integral, we find
\bea
   \nonumber
  \int d\text{LIPS}[\ell_1, \ell_2] \,(\bullet) &=&
  P^2 \int_{\tilde\lambda = \bar{\lambda}} \frac{\<\lambda, d\lambda \> [\tilde\lambda, d\tilde\lambda ]}{ \<\lambda | P | \tilde\lambda]^2} ~  \,g(\lambda)\\[2mm]
  &=&
  -  2\pi i \Big\{ g(\l_P)+P^2\sum\limits_{k}\frac{[k\,\h]}{\<k|P|k]\<k|
P|\h]} \Big(g{(\l)}\,\<k\,\l\>\Big)\Big|_{\l=k}\Big\}\,.~~~~
\label{resg} \eea where $\l_P=P|\h]$. The sum on the RHS of
\reef{lip1} runs over the simple poles of
$g(\l)$.\footnote{\label{footie:phi4}The manipulations carried out
here are also valid for 4-point 1-loop amplitudes in
$\Tr\,\phi^4$-theory. The scalars are taken to be in the adjoint of
some gauge group so we can consider color-ordered amplitudes.
For  $\phi^4$-theory $g(\lambda)=1$, and then
\reef{resg} gives
$C_\text{bubble}^\text{(L,R)}=1$
as needed.
We used this to fix the
normalization \reef{bubble1}.}

We can now evaluate the $d$LIPS integral in \reef{bubble2} to find the value of the bubble coefficient. We use $\widehat{\cs}^\text{\,[Cut (b)]}_{n,ij}  \Big|_{O(1)}$ from  \reef{1lp} in place of $(\bullet)$ in \reef{resg}, and the result is that in pure $\cn=1,2$ SYM the bubble coefficients are
\begin{equation}
\label{lip1}
C_\text{bubble}^\text{(L,R)} ~=~
- (4-\cn)\,\ca_{n,ij}^\text{tree}~
\bigg[ g(\l_P)+P^2\sum\limits_{k}\frac{[k\,\h]}{\<k|P|k]\<k|
P|\h]} \Big(g{(\l)}\,\<k\,\l\>\Big)\Big|_{\l=k} \bigg]  \, ,
\end{equation}
with
\begin{equation}\label{1lpm}
 g(\l) ~=~
 \frac{\<s-1,s\>\<r-1,r\>}{\<ij\>^2}
  \frac{\<i \lambda\>^2 \<j \lambda\>^2}
  {\<r-1\, \lambda\>\<r\, \lambda\>\<s-1\, \lambda\>\<s\, \lambda\>} \, .
\end{equation}
The result \reef{lip1} makes it very easy to obtain the bubble coefficients in pure $\cn=1,2$ SYM.

\vspace{2mm}
We now focus on the 4-point bubbles in  pure $\cn=1,2$ SYM. We have to consider two cases, depending on whether the external $\Psi$-states are adjacent or non-adjacent.

\vspace{4mm}
\noindent {\bf Adjacent case $\<\F_1\F_2\Psi_3\Psi_4\>$}\\
There is only one cut that separates the external $\Psi$-states, namely the 23-channel cut,
and this corresponds to $r=2$ and $s=4$ in \reef{1lpm}. Here $i=3$ and $j=4$, so we have
\begin{equation}
g({\l})=\frac{\<34\>\<12\>}{\<34\>^2} \frac{\<3\l\>\<4\l\>}{\<1\l\>\<2\l\>} \, .
\end{equation}
We choose $\eta=1$ in \reef{lip1}. Since $P=2+3$, we have
$\l_P = (p_2+p_3)|1] = -|4\>[41]$, so $g({\l_P})=0$. The sum in  \reef{lip1} is over $k = 1,2$, but the summand vanishes for $k=1$ because $\eta=1$. Hence the only non-vanishing contribution is from $k=2$ and it gives
\begin{eqnarray*}
C_\text{bubble}^{(23,41)}
&=& -(4-\cn)\,\ca_{4,34}^\text{tree}~
\<23\>[23]\,\frac{\<34\>\<12\>}{\<34\>^2}\frac{[21]\<32\>\<42\>}{\<2|3|2]\<2|3|1]\<12\>}
~=~-(4-\cn)\,\ca_{4,34}^\text{tree}\, .
\end{eqnarray*}

\vspace{4mm}
\noindent {\bf Non-adjacent case $\<\F_1\Psi_2\F_3\Psi_4\>$}\\
Now $i=2$ and $j=4$, so there are two contributing cuts, namely
\begin{enumerate}
\item 23-channel:~~  $r=2,~s=4$;~~$P=2+3$\,,
\item 12-channel:~~ $r=1,~s=3$;~~$P=1+2$\,,
\end{enumerate}
Eq.~\reef{1lpm} gives
\begin{equation}
g_1({\l})=\frac{\<12\>\<34\>}{\<24\>^2}\frac{\<2\l\>\<4\l\>}{\<1\l\>\<3\l\>}\,,
\hspace{8mm}
g_2({\l})=\frac{\<23\>\<41\>}{\<24\>^2}\frac{\<2\l\>\<4\l\>}{\<1\l\>\<3\l\>}\,.
\end{equation}
Use \reef{lip1} with $\eta=1$ to find
\bea
C_\text{bubble}^{(23,41)}=-(4-\cn)\,\ca_{4,24}^\text{tree}~\frac{\<12\>\<34\>}{\<13\>\<24\>}\,,
\hspace{1cm}
C_\text{bubble}^{(12,34)}=-(4-\cn)\,\ca_{4,24}^\text{tree}~\frac{\<14\>\<23\>}{\<13\>\<24\>}
\eea
Note that their sum is $C_\text{bubble}^{(23,41)} + C_\text{bubble}^{(12,34)} = -(4-\cn)\,\ca_{4,24}^\text{tree}$.


\subsection{Bubbles and the 1-loop $\beta$-function coefficient}
\label{s:beta}

The bubble contribution to the 1-loop amplitudes is
$\sum C_\text{bubble}^{i} I_\text{bubble}^{i}$. At leading order in dimensional regularization, the bubble integral is
\bea
  I_\text{bubble}^{i} = \frac{1}{(4\pi)^2} \frac{1}{\epsilon} + O(1) \, .
\eea
The coefficient of the $1/\epsilon$ term in the amplitude is thus the sum of the bubble coefficients.\footnote{The amplitude also has $(\log s_I)/\epsilon$ terms arising from the expansion of the soft IR divergences $(-s_I)^{-\epsilon}/\epsilon^2$, where $s_I$ are Mandelstam variables. These do not interfere with the $1/\epsilon$-terms discussed here.}
For 1-loop 4-point superamplitudes $A_{n,ij}^\text{1-loop}$ in pure $\cn=1,2$ SYM, we found above that
\bea
  \label{sumbubble}
  \sum C_\text{bubble} ~=~ -(\cn-4)\,\ca_{4,ij}^\text{tree} ~=~  - \beta_0\,\ca_{4,ij}^\text{tree}\, .
\eea
Here we have introduced the 1-loop $\beta$-function coefficient  $\beta_0$ defined by
\bea
  \mu \frac{dg}{d\mu} =\beta(g) = - \frac{\beta_0}{(4\pi)^2} g^3 + \dots
\eea
For pure $\cn=0,1,2$ SYM, $\beta_0 = 11/3, \,3$ and $2$, respectively.
It was shown in the \cite{ArkaniHamed:2008gz} that the result
$\sum C_\text{bubble} ~=~  - \beta_0\,A_{4,ij}^\text{tree}$ also holds
for 4-point amplitudes in  $\cn=0$ SYM.

The minus sign in  \reef{sumbubble} arises as follows. The bubble contribution $\frac{1}{(4\pi)^2 \epsilon}  \sum C_\text{bubble}$ does not capture the full UV divergence: it misses the UV divergences from bubbles on the external lines. In dimensional regularization, the UV divergences of bubbles on the external lines are precisely canceled by the collinear IR divergences. Thus
\be
  \nonumber
  \text{UV-div.}
  =
  \big(\sum C_\text{bubble} I_\text{bubble}\big)_\text{UV} 
  + \text{UV}_\text{ext.\,bubbles}
  = \big(\sum C_\text{bubble} I_\text{bubble}\big)_\text{UV} 
  - \text{IR}_\text{collinear}\,.~
\ee
 For an $n$-gluon 1-loop amplitude the collinear IR divergences take the form \cite{divs} 
\bea
  \label{coll}
   \text{$\text{IR}_\text{collinear}$:}~~~~~~~~~
   A^\text{1-loop}_{n,\text{collinear}}
  ~=~ -\frac{g^2}{(4\pi)^2} \frac{1}{\epsilon} ~ \frac{n}{2}\, \beta_0 A^\text{tree}_{n} \, .
\eea
At leading order in $\epsilon \to 0$,
the UV divergence is \cite{divs} 
\bea
  \label{UV}
   \text{$\text{UV-div.}$:}~~~~~~~~~
 A^\text{1-loop}_{n,\text{UV}}
  ~=~+\frac{g^2}{(4\pi)^2} \frac{1}{\epsilon} ~ \big(\tfrac{n}{2}-1\big)\, \beta_0 A^\text{tree}_{n} \, .
\eea
At MHV level, these relations generalize to superamplitudes in pure $\cn=1,2$ SYM, and adding \reef{coll} and \reef{UV} we have 
\bea
  \label{sumbubble2}
  \sum C_\text{bubble} = -\beta_0\,\ca_{n,ij}^\text{tree}
\eea
for all $n$.  
It is quite non-trivial from the point of view of the on-shell cut-construction of the bubble coefficients that \reef{sumbubble2} should hold.  It was established in \cite{GenUni} for bubbles of MHV amplitudes with an $\mathcal{N}=1$ chiral multiplet in the loop, and for  4-point  amplitudes in
pure YM theory in \cite{ArkaniHamed:2008gz} and YM theory with matter in   \cite{Lance}. Here we have 
verified the result \reef{sumbubble2} for 4-point amplitudes of pure $\cn=1,2$ SYM in a manifestly supersymmetric way using the on-shell superfield formalism.


\section{Solution to the SUSY Ward identities in $\cn<4$ SYM}\label{sec:SWI}

It has recently been shown \cite{Elvang:2009wd} that the on-shell SUSY Ward identities in $\cn=4$ SYM have a simple solution which presents the N$^K$MHV superamplitude  as a sum of SUSY and R-symmetry invariant Grassmann polynomials. Each invariant polynomial is multiplied by a basis amplitude; the number of algebraically independent basis amplitudes needed to determine an N$^K$MHV superamplitude is given by the dimension of the irrep of $SU(n-4)$ corresponding to the rectangular $4$-by-$K$ Young diagram. Moreover, the basis amplitudes are characterized precisely by the semi-standard tableaux of this Young diagram. The solutions to the SUSY Ward identities in $\cn=4$ SYM and $\cn=8$ supergravity and their applications are reviewed \cite{Elvang:2010xn}. 
In this section, we show that the solution from maximally supersymmetric Yang-Mills theory is easily generalized 
to $\cn<4$ SYM. 

At the level of superamplitudes, the SUSY Ward identities are equivalent to the statement that the SUSY charges ($\epsilon$ denote arbitrary Grassmann-odd spinors)
\bea
  Q^a = \sum_{i=1}^n [\e\, i] \frac{\partial}{\partial \eta_{ia}}\,,
  ~~~~~
  \tilde{Q}_a = \sum_{i=1}^n \<\e\, i\> \,\eta_{ia}\,,
  ~~~~~
  \text{for } a=1,2,\dots,\cn
\eea
annihilate the superamplitude, $Q^a \cf = \tilde{Q}_a \cf = 0$. Here we have specialized to the $\Phi$-$\Psi$ formulation of the on-shell superspace.
Both constraints are solved by the $\delta^{(2\cn)}$-function, 
provided momentum conservation is enforced, so the MHV superamplitudes $\cf_{n,ij}^\cn$ in \reef{MHVgenN} are manifestly supersymmetric.
The N$^K$MHV superamplitudes have Grassmann degree $\cn(K+2)$, so if they are written with an overall factor of $\delta^{(2\cn)}$, then the $\tilde{Q}_a$ SUSY Ward identities are satisfied, and one must then just ensure that the order $\cn K$ polynomial multiplying $\delta^{(2\cn)}$ is annihilated by $Q^a$.

Rather than deriving the most general solution, we simply illustrate the procedure in the simple case of the NMHV sector of $\cn=1$ SYM.
Let lines $u$, $v$ and $w$ be the $\Psi$-sector states.
We then write the NMHV superamplitude of $\cn=1$ SYM as
\bea
  \label{Fwi1}
  \cf_{n,uvw}~=~
  \delta^{(2)}(\tQ)
  \sum_{i=1}^n f_i\,\eta_{i}
  ~=~
  -\frac{1}{\<vw\>}  \delta^{(2)}(\tQ)
  \sum_{i\ne v,w} c_i\,\eta_{i}\,.
\eea
In the second equality we have used the $\delta$-function to eliminate $\eta_v$ and $\eta_w$ from the sum and included a convenient normalization factor $-\<vw\>^{-1}$. The coefficients $c_i$ can be written in terms of the $f_i$, but their specific relationship is not needed in the following.

The requirement $Q^a \cf_{n,uvw}=0$ now turns into the condition
\bea
   \sum_{i\ne v,w}  [\e\, i] \,c_i\, = 0
   ~~~~\longrightarrow~~~~
   \left\{
   \begin{array}{l}
   [r\,s]\,c_s + \sum_{i\ne v,w,r,s}  [r\, i] \,c_i\, = 0\\[2mm] {}
   [s\,r]\,c_r + \sum_{i\ne v,w,r,s}  [s\, i] \,c_i\, = 0
  \end{array}
   \right.
\eea
We have selected two lines $r,s \ne u,v,w$, and used $\e=r,s$ to extract the two conditions that are now used to eliminate $c_r$  and $c_s$ from \reef{Fwi1}. The result is
\bea
  \label{Fwi2}
  \cf_{n,uvw}
  ~=~
  -\frac{1}{\<vw\> [rs]}
  \delta^{(2)}(\tQ)
  \sum_{i\ne v,w} c_i\,m_{rsi}
\eea
where
\bea
   m_{rsi} = [rs]\, \eta_i + [s\,i]\, \eta_r + [i\,r]\, \eta_s \, .
\eea
Note that $\tQ\, m_{rsi} = 0$ thanks to the Schouten identity. The polynomial $m_{rsi}$ is familiar from the $3$-point anti-MHV superamplitudes.

Now the final step is to identify the $c_i$ as basis amplitudes for the superamplitude $\cf_{n,uvw}$. Let us project out negative helicity gluons on lines $v$ and $w$; this amounts to applying $-\partial_v \partial_w$ to  $\cf_{n,uvw}$. The derivatives only hit $\delta^{(2)}$ and the result is a factor $-\<vw\>$ that cancels the same factor in the denominator in \reef{Fwi2}. We need to apply one more $\partial$ to extract a component amplitude. There are two options: 1) applying $\partial_u$ is equivalent to taking state $u$ to be a negative helicity gluon. The derivative produces a factor $[rs]$ from $m_{rsu}$
so the result is
\bea
  \label{cu}
  c_u ~\sim~\< \ldots -_u \ldots -_v \ldots -_w \ldots\>
\eea
where dots ``\ldots''  stand for positive helicity gluons. The other option 
is 2)
 applying $\partial_k$ for $k\ne u,v,w,r,s$. This designates $k$ as a positive helicity gluino and forces $u$ to be negative helicity gluino; hence
\bea
  \label{ck}
  c_k ~\sim~\< \lambda_k^+ \ldots \lambda^-_u \ldots -_v \ldots -_w \ldots\>
  ~~~~\text{for } k\ne u,v,w,r,s
\eea
We use $\sim$ here to indicate that minus signs arise when the derivative $\partial_k$ is required to move past an odd number of $\Psi$ states. Also, the position of $\lambda_k^+$ is only indicated schematically and depends on the value of  $k$ relative to $u$, $v$, $w$.

With $c_i$'s identified in \reef{cu} and \reef{ck},
the result \reef{Fwi2} is then our manifestly supersymmetric $\cn=1$ NMHV superamplitude.  The basis amplitudes are the $n\!-\!5$ gluino 
amplitudes \reef{ck} and the pure gluon amplitude \reef{cu}. This is a total of $n\!-\!4$ basis amplitudes. 
For $n=6$ this is the familiar result of \cite{Bianchi:2008pu,Grisaru:1977px}  that 2 basis amplitudes are required to determine all amplitudes in each of the 3 NMHV sectors. Our basis here is different from that of \cite{Bianchi:2008pu,Grisaru:1977px}; we made choices above that fixed our basis. For example, we selected to eliminate $\eta_v$ and $\eta_w$ and this fixed the states $v$ and $w$ to be negative helicity gluons. If we had chosen to eliminate the $\eta$ of a $\Phi$ state instead, then that line would have been fixed to be a positive helicity gluino. The choices that lead to \reef{Fwi2} are equivalent to those made in the $\cn=4$ SYM analysis of \cite{Elvang:2009wd,Elvang:2010xn}, so indeed we could just have carried out the truncation procedure of the $\cn=4$ result. We found it useful to carry out the analysis here to illustrate it in  the   
much simpler context of $\cn=1$ SYM.

Going beyond NMHV is easy in $\cn=1$ SYM. Now one needs a polynomial
$\sum c_{\{i_k\}} \eta_{i_1} \dots \eta_{i_K}$. The coefficients $c_{\{i_k\}}$ are fully anti-symmetric in the indices $\{i_k\}$. As above, $\tQ$-SUSY allow us to fix two $\Psi$ states to be negative helicity gluons and $Q$-SUSY can fix two $\Phi$-states to be positive helicity gluons. The remaining $n\!-\!4$ states are $K$ $\Psi$ states and $n\!-\!4\!-\!K$ $\Phi$ states. The algebraic basis consists of amplitudes with $m$ pairs $\l^+ \l^-$ on the $n\!-\!4$ unfixed lines. There are $\binom{K}{m}$ ways to choose the position of the $m$ $\l^-$'s and $\binom{n-4-K}{m}$ ways to choose the position of the $m$ $\l^+$'s. $m=0$ is the unique\footnote{Recall that the $\Psi$-states are fixed.} pure gluon amplitude and we can maximally have $K$ pairs $\l^+ \l^-$; hence the total number of N$^K$MHV basis amplitudes is
\bea
  \sum_{m=0}^K  \binom{K}{m} \binom{n-4-K}{m}
  =
  \binom{n-4}{K}\,.
\eea
This number is also the dimension of the fully anti-symmetric irrep of $SU(n-4)$, whose Young diagram is rectangular $1$-by-$K$.

For $\cn=2$ the analysis can be carried out similarly, now also incorporating the $SU(2)$ R-symmetry. For $\cn=3$ the analysis is completely analogue, and once all possible positions of the $\Psi$-states are considered the result should be  equivalent to the $\cn=4$ SYM result.

\section{Spinor helicity and SYM amplitude in 6d}
\label{sec:6d}

The recently developed 6d spinor helicity formalism \cite{6Dspinor1,6Dspinor2}  can be combined with 6d on-shell superfield formalism to encode amplitudes of 6d pure SYM into superamplitudes \cite{6Dspinor3}. Breaking the manifest 6d Lorentz invariance allows us to find a direct connection between the 6d and 4d superamplitudes. More precisely, the amplitudes of $\cn=4$ SYM away from the origin of moduli space (Coulomb branch) are obtained by interpreting the extra components of the 6d momenta as 4d masses. This approach has been utilized to obtain the massively regulated 4d $\cn=4$ SYM loop amplitude from that of the 6d $\cn=(1,1)$ SYM theory~\cite{6loopDamplitudeYT,6loopDamplitudeQM}. Setting the extra components to zero one obtains the 4d massless amplitude in a non-chiral formulation. In this section we will demonstrate that the non-chiral formulation in 4d 
implies non-trivial relations between N$^K$MHV amplitudes of different $K$. Furthermore, we can truncate the 6d theory to $\cn=(1,0)$ SYM, and from this obtain the $\cn=2$ SYM amplitudes in 4d. We provide the detailed connection between the 6d and 4d superspace in this section.

\subsection{Spinor helicity in 6d and maximal SYM}

We briefly outline needed aspects of the 6d spinor helicity formalism \cite{6Dspinor1,6Dspinor2}.
In six dimensions, the massless Dirac equation reads
\bea
   p_{AB} \lambda^{B a} = 0\,,~~~~~
   p^{AB} \tilde{\lambda}_{B\dot{a}} = 0\,,~~~~~
\eea
with $p_{AB} \equiv p_\mu \sigma^\mu_{AB}$ and $p^{AB} \equiv p^\mu \tilde{\sigma}_{\mu}^{AB}$. The $\sigma$'s are the 6d Pauli matrices with indices $A,B,..=1,2,3,4$ of $Spin(1,5)\sim SU^*(4)$.\footnote{Here the $*$ means it is pseudoreal, where the reality condition is defined using the $SU(2)$ little group, i.e. they are ``$SU(2)$-Majorana" spinors~\cite{Kugo:1982bn}. } The Pauli matrices are chosen such that they coincide with our usual 4d $\g$-matrices for $\mu=0,1,2,3$, and an explicit form can be found in~\cite{6Dspinor1}.
The two representations are related by
$p_{iAB}=\frac{1}{2}\epsilon_{ABCD}p_i^{CD}$, and the null vector condition is simply
$p^{AB}p_{AB} \propto  p^2= 0$.
The Dirac equation has two independent solutions labelled by indices
$a=1,2$ and $\dot{a}=1,2$ of the little group $SO(4) \sim SU(2) \times SU(2)$.\footnote{The little group indices are raised and lowered by $\e_{ab}$ and $\e^{\dot{a}\dot{b}}$ as $\lambda_a = \e_{ab}\lambda^b$ and
$\tilde\lambda^{\dot{a}} = \e^{\dot{a}\dot{b}}\tilde\lambda_{\dot{b}}$
with $\epsilon_{12}=-1$, $\epsilon^{12}=1$.}
The null momentum can be expressed as bi-spinors, viz.
\bea
p_i^{AB}=\lambda_i^{A a}\,\epsilon_{ab}\lambda_i^{B b} \,,
\hspace{1cm}
p_{iAB}=\tilde{\lambda}_{iA\dot{a}}\,\epsilon^{\dot{a}\dot{b}}\tilde{\lambda}_{iB\dot{b}}
\,.
\label{VectorSpinor}
\eea
Further details of the six-dimensional spinor helicity formalism can be found in ref.~\cite{6loopDamplitudeYT,6Dspinor1}.

Maximal super Yang-Mills in 6d has $\cn=(1,1)$ supersymmetry. The on-shell supermultiplet contains gluons with four polarization states
$g^a\,_{\dot{a}}$, four scalars $(\phi,\phi',\phi'',\phi''')$, and
eight fermions
$(\psi_{\dot{a}},\tilde{\psi}_{\dot{a}},\chi_a,\tilde{\chi}_a)$. These states can be encoded compactly in a single superwavefunction using two sets of anticommuting Grassmann variables $\xi_{a}$ and $\tilde{\xi}^{\dot{a}}$ which carry the little group indices. As opposed to their 4d equivalents, $\xi_a$ and $\tilde{\xi}^{\dot{a}}$ do not carry $SU(2)\times SU(2)$ R-symmetry indices; they are chosen to make the little group manifest  instead of the R-symmetry \cite{6Dspinor3}. The superwavefunction takes the form
\begin{eqnarray}
\Phi_\text{6d}(\xi,\tilde{\xi}) &=&
  \phi
  + \chi^a \xi_a
  + \phi'(\xi)^2
  + \psi_{\dot{a}}\tilde{\xi}^{\dot{a}}
  + g^a\,_{\dot{a}}\xi_a\tilde{\xi}^{\dot{a}}
  + \tilde{\psi}_{\dot{a}}(\xi)^2\tilde{\xi}^{\dot{a}} \nonumber\\[2mm]
  && \null
  + \phi''(\tilde{\xi})^2
  + \tilde{\chi}^a\xi_a(\tilde{\xi})^2
  + \phi'''(\xi)^2(\tilde{\xi})^2 \,,
  \label{expansion}
\end{eqnarray}
where $\xi^2=\frac{1}{2}\xi^a\xi_a=\xi_2\xi_1$ and
$\tilde{\xi}^2=\frac{1}{2}\tilde{\xi}_{\dot{a}}\tilde{\xi}^{\dot{a}}=
\tilde{\xi}^{\dot{1}}\tilde{\xi}^{\dot2}$. We raise or lower
the indices as $\x^a=\e^{ab}\x_b$ and
$\tilde{\x}_{\dot a}=\e_{\dot a\dot{b}}\x^{\dot{b}}$.

The SUSY charges, or supermomenta, take the form
\eq
q_i^A=\lambda_i^{Aa}\xi_{ia},\;\;~~~~~~
\tilde{q}_{iA}=\tilde{\lambda}_{iA\dot{a}}\tilde{\xi}_{i}^{\dot{a}}\,.
\eqe
The four- and five-point amplitudes are then given by ($\delta^6(\sum p)$ is implicit)
\eqa
\label{6d4ptmax}
\cn=(1,1):\;\;\mathcal{M}_{4}&=&
-\frac{i\,\delta^{(4)}(\sum q)\,\delta^{(4)}(\sum \tilde{q})}{st}\\[3mm]
\cn=(1,1):~~\mathcal{M}_5&=&
\;\frac{i\delta^{(4)}(\sum q)\,\delta^{(4)}(\sum\tilde{q})}{s_{12}s_{23}s_{34}s_{45}s_{51}}
\bigg\{q_{1}(\displaystyle{\not}p_2 \displaystyle{\not}p_3\displaystyle{\not} p_4\displaystyle{\not} p_5)\tilde{q}_{1}+\text{cyclic}\;\big.
\nonumber
\\
\label{6dmaxamp}
&&\hspace{3.8cm}
 +\frac{1}{2}\Big[
   \Delta_{12} +  \Delta_{34} +  \Delta_{45} +  \Delta_{35}+ \text{c.c.}
 \Big] \bigg\}\,,~~~~~~~
\eqae
where $\delta^{(4)}(\sum q)=\frac{1}{4!}\epsilon_{ABCD}(\sum q^A)(\sum q^B)(\sum q^C)(\sum q^D)$, and similarly for $\tilde{q}$. For $n=5$ we have introduced
$\Delta_{ij} \equiv  q_i ( \displaystyle{\not}p_{j~} \displaystyle{\not}p_{j+1\,} \displaystyle{\not}p_{j+2\,} \displaystyle{\not}p_{j+3\,}
- \displaystyle{\not}p_{j~} \displaystyle{\not}p_{j+3\,} \displaystyle{\not}p_{j+2\,} \displaystyle{\not}p_{j+1\,}) \tilde{q}_j$. The 3-point superamplitudes require  additional bosonic variables due to the special kinematics~\cite{6Dspinor1,6loopDamplitudeYT}.

For higher point amplitudes, we simply note the structure of the supermomentum $q,\tilde{q}$. The Grassmann degree of the $n$-point amplitude can be deduced by the requirement of R-invariance. The generators of the $U(1)\times U(1)$ subgroup of $SU(2)\times SU(2)$  R-symmetry takes the form
\eq
J^{U(1)}=\sum_i \left(\xi_{ia}\frac{\partial}{\partial \xi_{ia}}-1\right)\,,\;~~~~~\tilde{J}^{U(1)}=\sum_i \left(\tilde{\xi}_{i\dot{a}}\frac{\partial}{\partial \tilde{\xi}_{i\dot{a}}}-1\right)\,.
\eqe
These generators can be derived by considering the twistor representation of the $\mathcal{N}=(1,0)$ superconformal group OSp$^*$(8$|$2).\footnote{Again, the $*$ here indicates pseudoreality.} The constant piece in the generators can be checked by anticommuting the supersymmetry and conformal supersymmetry generators; details of these generators are given in \cite{6D20}. R-invariance of the $n$-point amplitude requires it to be of degree $(n,n)$ in $(q,\tilde{q})$.
Four $q$'s and four $\tilde q$'s are accounted for by the supermomentum delta functions, so the $n$-point amplitude will be proportional to a polynomial of degree $q^{n-4}\tilde{q}^{n-4}$. Lorentz and little group invariance require that these $q^{n-4}\tilde{q}^{n-4}$ will appear in the amplitudes as products of
\eq
\begin{array}{cc}\text{odd} & \text{even} \\q\;\overbrace{\displaystyle{\not}p_i\cdot\cdot\displaystyle{\not}p_k} q\;,& q\;\overbrace{\displaystyle{\not}p_i\cdot\cdot\displaystyle{\not}p_k}\tilde{q}\end{array}\,.
\label{qcontraction}
\eqe
The contractions of supermomenta of the same chirality requires odd power of momenta, while for opposite chirality an even number of momenta is needed.
For example, a BCFW construction \cite{6loopDamplitudeYT} indicates that the 6-point superamplitude includes terms of the form
\eqa
\mathcal{S}(p)\left(q\;\displaystyle{\not}\mathcal{P}_\text{odd} q\right)\left(\tilde{q}\;\displaystyle{\not}\mathcal{P}_\text{odd} \tilde{q}\right),
\hspace{1cm}
\tilde{\mathcal{S}}(p)\left(\tilde{q}\;\displaystyle{\not}\mathcal{P}_\text{even} q\right)\left(\tilde{q}\;\displaystyle{\not}\mathcal{P}_\text{even} q\right)
\label{6pointq}
\eqae
where $\displaystyle{\not}\mathcal{P}_\text{odd/even}$ represents strings of even (including zero), or odd number of momenta, while $\mathcal{S}$ and $\tilde{\mathcal{S}}$ represents pure momentum inner products.

\subsection{4d-6d correspondence}
Details of the reduction of component amplitudes from 6d to 4d can be found in ref.~\cite{6Dspinor1,6loopDamplitudeYT}; here we restrict ourselves to the massless case. The massless 4d amplitudes are obtained by restricting the 6d momenta to the 4d subspace with $p_4=p_5=0$. The 6d Dirac spinors are then simply given in terms of the 4d massless spinors. Written as $4 \times 2$ and $2 \times 4$ matrices, they take the form
\eqa
\lambda^A_{ia}=\left(\begin{array}{cc}0& |i\rangle
\\ |i]
&0\end{array}\right),\;\;
~~~~~
\tilde{\lambda}_{iA\dot{a}}=\left(\begin{array}{cc}0
&\langle i| \\ -[i|
&0\end{array}\right).
\label{cusp}
\eqae

\begin{figure}
\begin{center}
\includegraphics[scale=0.9]{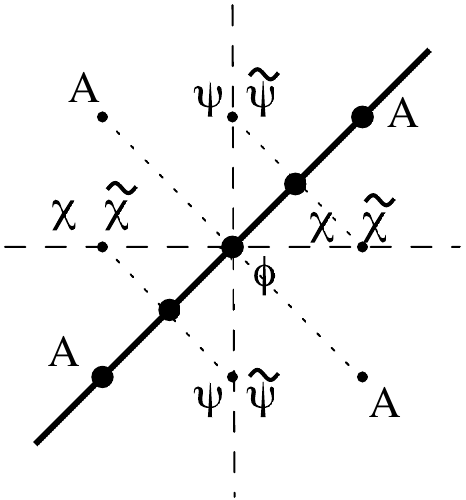}
\caption{\small The projection of the 6d $\mathcal{N}=(1,1)$ multiplet onto the 4d $\mathcal{N}=4$ multiplet. The two axes are the weights of the states with respect to the two $U(1)$'s of little group $SU(2) \times SU(2)$. The diagonal line represents the $U(1)$ subgroup of the 4d little group.}
\label{projection}
\end{center}
\end{figure}

The reduction of the on-shell 6d superspace variables $\xi^a$ and $\tilde\xi_{\dot{a}}$ gives a non-chiral representation of the 4d superspace involving both $\eta$ and $\bar{\eta}$. A specific choice is
\eqa
\nonumber D=6:\;\left(\begin{array}{c}\xi_1 \\[1mm] \xi_2\end{array}\right) &\rightarrow& \;D=4:\; \left(\begin{array}{c}\eta_1 \\[1mm] \overline{\eta}^2 \end{array}\right)\\[2mm]
D=6:\;\left(\begin{array}{c}\tilde{\xi}^{\dot{1}} \\[1mm] \tilde{\xi}^{\dot{2}}\end{array}\right) &\rightarrow& \;
D=4:\; \left(\begin{array}{c} \overline{\eta}^3 \\[1mm] \eta_4\end{array}\right)\,.
\label{etaID} \eqae
Writing the 6d superwavefunction $\Phi_\text{6d}$ of \reef{expansion} in this form makes it possible to directly compare with the 4d superwavefunction $\Omega_\text{4d}$ in \reef{n4Phi}. All that is needed is a Fourier transform of the variables $\eta_2$ and $\eta_3$; this gives\footnote{This on-shell superfield is closely related to the scalar superfield in projective superspace~\cite{Hatsuda:2008pm}.}
 \bea
  \nonumber
  \Omega'_\text{4d}
  &=& \int d\eta_{2}\, d\eta_{3}~e^{\eta_2 \bar{\eta}^2}\,
  e^{\eta_3 \bar{\eta}^3}~ \Omega_\text{4d}
  \\[1mm]
  \nonumber
  &=&
  S^{23} - ( \lambda^{123} \eta_1 +  \lambda^3 \bar\eta^2)
  + ( \lambda^2 \bar\eta^3- \lambda^{234}\eta_4 )
  +\bar\eta^2\eta_1  S^{13}
  +\bar\eta^3\eta_4  S^{24}
  \\[1mm]
  \nonumber
  &&
  - \eta_1 \eta_4 G^-
  + \bar\eta^2 \bar\eta^3 G^+
  +\eta_1 \bar\eta^3 S^{12}
  - \bar\eta^2 \eta_4 S^{34}
  \\[1mm]
  &&
  -\bar\eta^3  \eta_4 ( \lambda^{124} \eta_1 + \lambda^4  \bar\eta^2)
  +\bar\eta^2 \eta_1  (  \lambda^1 \bar\eta^3 - \lambda^{134} \eta_4)
  +  \bar\eta^2 \eta_1 \bar\eta^3 \eta_4  S^{14} \, .
\eea
Comparing $\Omega'_\text{4d}$ with \reef{expansion}, using the identification \reef{etaID}, we can now identify the on-shell states of $\cn=(1,1)$ SYM in 6d with the massless on-shell states of $\cn=4$ SYM in 4d:
\bea
  \nonumber
  \text{Scalars:}  &&
  \phi = S^{23}\,,~~~
  \phi' = S^{13}\,,~~~
  \phi'' = S^{24}\,,~~~
  \phi''' = S^{14}\,, \\[3mm]
  \nonumber
  \text{Gluinos:}  &&
  \chi^a = -(\lambda^{123},\lambda^3)\,,~~~~~
  \psi_{\dot{a}} = (\lambda^2,-\lambda^{234})\,,\\[1mm]
  \nonumber
  &&
  \tilde\chi^a = -(\lambda^{124},\lambda^4)\,,~~~~~
  \tilde\psi_{\dot{a}} = (\lambda^1, -\lambda^{134},)\,,
  \\[2mm]
  \text{Gluons:} &&
  g^{a}_{~\;\dot{a}}
  =
  \left(
  \ba{cc}
  S^{12} & -G^-  \\
  G^+ & - S^{34}
  \ea
  \right)
 \label{componentID}\,.
 \eea
In figure \ref{projection} this identification is illustrated very intuitively as a projection of the $\mathcal{N}=(1,1)$ multiplet onto $\mathcal{N}=4$ multiplet.

We have identified the states and superwavefunctions in the reduction of 6d $\cn=(1,1)$ SYM to 4d $\cn=4$ SYM. The 6d 4-point superamplitudes are also mapped the 4d ones. Using \reef{cusp} and \reef{etaID}, the 6d supermomenta take the form
\eq
{q}_{i}^A\rightarrow \left(\begin{array}{c}
|i\rangle \,\eta_{i1} \\[2mm]
\!-|i] \,\bar{\eta}_i^2 \end{array}\right) \, ,
 ~~~~~~~~
 \tilde{q}_{iA}\rightarrow \Big(\,- [ i|\, \bar{\eta}_i^3\;, \langle i|\, \eta_{i4}\Big)\,.
 \;\;
\label{qreduc}
\eqe
The supermomentum delta functions defined below \reef{6dmaxamp} can then be written in terms of 4d variables as
\eqa
\nonumber \delta^4\Big(\sum_i q_i^A\Big)&=&
\Big(\sum_{i,j}\eta_{i1}\langle ij\rangle\eta_{1j}\Big)\Big(\sum_{l, k}\bar{\eta}^2_{l}\left[ lk\right]\bar{\eta}^2_{k}\Big) \, ,\\
 \delta^4\Big(\sum_i \tilde{q}_{iA}\Big)&=&
 \Big(\sum_{i, j}\eta_{i4}\langle ij\rangle\eta_{4j}\Big)\Big(\sum_{l, k}\bar{\eta}^3_{l}\left[ lk\right]\bar{\eta}^3_{k}\Big) \, .
 \label{6ddelta}
\eqae
Applying this to the 4-point superamplitude \reef{6d4ptmax} gives an unfamiliar non-chiral  form of the 4d Parke-Taylor superamplitude. We can recover the more familiar chiral form with $\delta^{(8)}$ by performing a half-Fourier transformation. The details can be found in section 6.2 of \cite{Hatsuda:2008pm}.

\subsection{4d N$^K$MHV helicity sectors from 6d}
It is easy to track how the N$^K$MHV 
helicity sectors arise in the 6d-4d reduction using superamplitudes.  Let us start with the 5-point amplitude \reef{6dmaxamp} in 6d. The reduction to 4d yields  two different 4d structures, namely
\eq
 \label{M5-6d4d}
\mathcal{A}_5\;\;\; ~~~~
\text{6d}:~~q(\displaystyle{\not}p\displaystyle{\not}p\displaystyle{\not} p\displaystyle{\not} p)\tilde{q}
~~~\rightarrow ~~~
\text{4d}:~~\left(\begin{array}{c} \eta_{i1}\langle i\left| \displaystyle{\not}p\displaystyle{\not}p\displaystyle{\not} p\displaystyle{\not} p\right|j\rangle \eta_{j4} \\[2mm]
 \bar{\eta}^2_i\left[ i\left|\displaystyle{\not}p\displaystyle{\not}p\displaystyle{\not} p\displaystyle{\not} p\right|j\right]\bar{\eta}_j^3 \end{array}\right)
\eqe
No $\eta\bar{\eta}$-terms appear since this would require odd number of momenta between the 4d supermomenta, and such terms do not appear in the 6d parent superamplitude. The delta-functions supply 4 $\eta$'s and 4 $\bar\eta$'s, so after performing the $2$ inverse Fourier transformations for each of the 5 external lines, we obtain $\eta$-polynomials of degrees 12 and 8 respectively from the two structures in \reef{M5-6d4d}. Thus the two different $\eta$-$\bar{\eta}$ structures in \reef{M5-6d4d} encode the 5-point anti-MHV and MHV sectors in 4d.

For the 6-point superamplitude, things become more interesting. A general simplified form of the 6-point superamplitude is not yet known, but BCFW indicates that the 6d supermomenta appear in the amplitude as in (\ref{6pointq}). After reduction to 4d, we have
\bea
 \label{6point1}
&&\begin{array}{ccccl}
\text{6d}:&
\left(q\;\displaystyle{\not}\mathcal{P}_\text{odd} q\right)
\left(\tilde{q}\;\displaystyle{\not}\mathcal{P}_\text{odd} \tilde{q}\right)
&~\rightarrow&
\text{4d}:&~~~~~~ \eta_{i} \eta_{j} \bar\eta_{k}  \bar\eta_{l}\,
\langle i | \displaystyle{\not}\mathcal{P}_\text{odd}|k]
\langle j | \displaystyle{\not}\mathcal{P}_\text{odd}|l]
\end{array}\hspace{2cm}
 \\[5mm]
  \label{M6-6d4d2}
&&\begin{array}{ccccc}
\text{6d}:&
\left(q\;\displaystyle{\not}\mathcal{P}_\text{even} \tilde{q}\right)
\left(q\;\displaystyle{\not}\mathcal{P}_\text{even} \tilde{q}\right)
&\rightarrow&
\text{4d}:&\left(\begin{array}{l}
\eta_{i} \eta_{j}\eta_{k} \eta_{l}  \,
\langle i|\;\displaystyle{\not}\mathcal{P}_\text{even}|j\>\,
\langle k|\;\displaystyle{\not}\mathcal{P}_\text{even}|l\> \\[2mm]
 \eta_{i} \eta_{j}\bar\eta_{k}\bar\eta_{l}  \,
\langle i|\;\displaystyle{\not}\mathcal{P}_\text{even}|j\>\,
[ k|\;\displaystyle{\not}\mathcal{P}_\text{even}|l] \\[2mm]
 \bar\eta_{i} \bar\eta_{j} \bar\eta_{k} \bar\eta_{l}  \,
[ i|\;\displaystyle{\not}\mathcal{P}_\text{even}|j]\,
[ k|\;\displaystyle{\not}\mathcal{P}_\text{even}|l]
 \end{array}\right)
\end{array}~~~~
\label{6point2}
\eea
There are three different $\eta$-$\bar\eta$ structures:
$\eta \eta \eta \eta$ corresponds to anti-MHV,
$\eta \eta \bar\eta \bar\eta$ to NMHV, and $\bar\eta \bar\eta \bar\eta \bar\eta$ to MHV. The (anti)MHV amplitudes only have contributions from structures with $\displaystyle{\not}\mathcal{P}_\text{even}$. The $\displaystyle{\not}\mathcal{P}_\text{odd}$ terms are needed to get the full NMHV answer, and clearly this information is not contained in the MHV or anti-MHV amplitudes.

\begin{figure}
\begin{center}
\centerline{\includegraphics[width=9cm]{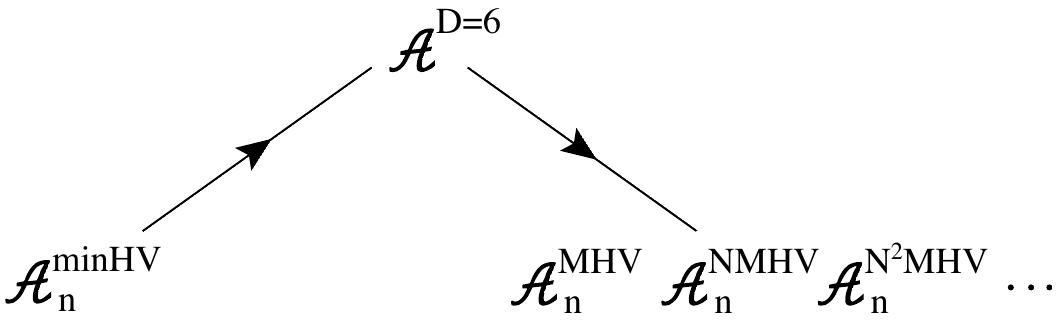}}
\begin{picture}(0,0)(0,0)
    \put(-10,14){?}
\end{picture}
\vspace{-5mm}
\caption{\small A proposed (not proven) scenario connecting minHV amplitudes to all N$^K$MHV amplitudes via the 6d reconstruction.}
\label{windy}
\end{center}
\end{figure}
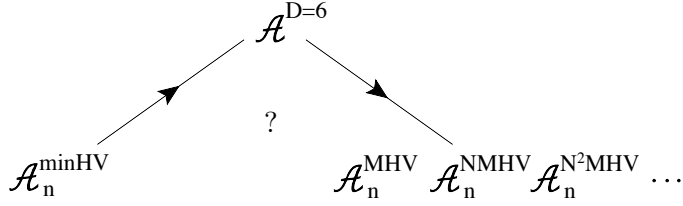

Note that the form of $\displaystyle{\not}\mathcal{P}_\text{even}$  is the same for each term in (\ref{6point1}). The same holds for $\displaystyle{\not}\mathcal{P}_\text{odd}$ in (\ref{6point2}). Since both (\ref{6point1}) and (\ref{6point2}) contributes to the NMHV amplitude, it appears that the NMHV amplitude is sufficient to capture the full structure of the 6d parent amplitude. At  higher $n$, one may then suspect that the ``minimally helicity violating" (minHV) amplitudes N$^{\lfloor\frac{n-4}{2}\rfloor}$MHV capture the complete structure of the parent 6d amplitude. IF true, this would give a relation between different helicity structures in 4d. Start with the 4d minHV amplitude and perform a half Fourier transformation to the non-chiral basis. From the parent 6d structure, we know there has to be a way to rewrite the spinor inner-products as spinor-traces of momenta or momentum inner-products.\footnote{This step is non-trivial and could be practically very challenging.} It would lead to a form of the superamplitude which depends only on momentum and supermomentum, cf.~\reef{qcontraction}. Lifting the minHV amplitudes to  6d allows one to complete the superamplitude because all $\displaystyle{\not}\mathcal{P}_\text{even/odd}$ structures are obtained from minHV. Reducing back to 4d one can now recover all other N$^K$MHV sectors. The procedure is illustrated in figure \ref{windy}.
From a practical view point this is not particularly useful since the minHV amplitudes are the most complicated helicity configuration for $n>5$. We mention it here only to illustrate the structure of the amplitudes; further evidence that the minHV determine all amplitudes would be desirable.

Note that the above statement is based on global symmetries of the amplitude, and hence should be valid for loop amplitudes as well. However a naive analysis of the six-point amplitude seems to contradict this statement. The known expression for the one-loop six point MHV amplitude includes the two-mass-easy box integrals, while the NMHV amplitude includes two-mass-hard integrals~\cite{GenUni,1L6P}. These two integrals are linearly independent and hence it is unlikely that NMHV amplitudes contains information of the MHV amplitude. The resolution is that the familiar 4d integral basis does not form an independent basis from the 6d point of view. For example the 5-point one loop amplitude in 6d is given in terms of scalar box plus pentagon integrals~\cite{6loopDamplitudeQM}. In the reduction to 4d, the 6d scalar pentagon reduces to five different scalar box integrals plus terms that vanish in 4d~\cite{Bern:1993kr}. Thus in lifting the NMHV loop amplitude to 6d, one is required to take into account such integral reduction identities to obtain the full 6d amplitude. An explicit six-point computation would be a first step to clarify these issues.

\subsection{$\mathcal{N}=(1,0)$ amplitudes from $\mathcal{N}=(1,1)$}
The six-dimensional $\mathcal{N}=(1,0)$ super Yang-Mills multiplet contains on-shell four gluon polarization states $g^a\,_{\dot{a}}$ and four chiral fermions $\psi_{\dot{a}},\tilde{\psi}_{\dot{a}}$. We have labelled the fields such that the embedding in the maximal multiplet is clear. The fields of the non-maximal multiplet are contained in two superfields ($\dot{a}=1,2$):
\eq
\;\;\Upsilon_{\dot{a}}(\xi)=\psi_{\dot{a}}+ g^a\,_{\dot{a}}\,\xi_a+\xi^2\,\tilde{\psi}_{\dot{a}}
\label{1,0}
\eqe
In the superamplitude, each external field can be assigned to different multiplet labelled by $\dot{a}$. In six dimensions, there are no selection rules since the continuous SU(2) little group will rotate between all possible assignments of $\dot{a}$.

{}From eq.~(\ref{1,0}) one can immediately read off the prescription of obtaining the two multiplets from the $\mathcal{N}=(2,0)$ multiplet, i.e. one simply integrates away one $\tilde{\xi}$ and set the other to zero. Schematically one has:
\eq
\Upsilon_{\dot{a}}(\xi)=\int d\tilde{\xi}^{\dot{a}}~\Phi_\text{6d}(\xi,\tilde{\xi})\Big|_{\tilde{\xi}=0}
\eqe
In terms of amplitudes, as with four-dimensons, one can start with the maximal $\mathcal{N}=(1,1)$ amplitudes, integrate out $\tilde{\xi}$ and one obtains the $\mathcal{N}=(1,0)$ amplitudes after setting the remaining $\tilde{\xi}$ to zero. The only difference now is the we need to integrate one $\tilde{\xi}$ for each external lines, compared to the four-dimensional case where we only integrate those lines that are negative helicity.

Integrating away the one $\bar{\xi}$ for each external leg and setting the remaining ones to zero, one see that the four- and five-point amplitudes are given by:
\eqa
\label{6d4ptN1}
\cn=(1,0)\!:~~~
\mathcal{M}_{4,\dot{a}\dot{b}\dot{c}\dot{d}}&=&
-\frac{i\delta^6(\sum p)\,\delta^4(\sum q)~
\langle 1_{\raisebox{-2pt}{${\scriptstyle\dot{a}}$}}2_{\dot{b}}3_{\raisebox{-2pt}{${\scriptstyle\dot{c}}$}}4_{\dot{d}}\rangle}{st}\\
\nonumber \cn=(1,0)\!:
~~\mathcal{M}_{5,\dot{a}\dot{b}\dot{c}\dot{d}\dot{e}}&=&
-\frac{i\delta^4(\sum q)}{s_{12}s_{23}s_{34}s_{45}s_{51}}
\Big\{q_{1}(\displaystyle{\not}p_2\displaystyle{\not} p_3 \displaystyle{\not}p_4 \displaystyle{\not}p_5)
\,\tilde{\lambda}_{1\dot{a}}
\,\langle 2_{\dot{b}}3_{\raisebox{-2pt}{${\scriptstyle\dot{c}}$}}4_{\dot{d}}5_{\raisebox{-2pt}{${\scriptstyle\dot{e}}$}}\rangle
+\text{cyclic}\;\big.\\
\label{N=10amps}
&&
+\frac{1}{2}\left[q_{1}(\displaystyle{\not}p_2\displaystyle{\not} p_3 \displaystyle{\not}p_4\displaystyle{\not} p_5
-
\displaystyle{\not}p_2 \displaystyle{\not}p_5\displaystyle{\not}p_4 \displaystyle{\not}p_3)
\,\tilde{\lambda}_{2\dot{b}}
\,\langle 3_{\raisebox{-2pt}{${\scriptstyle\dot{c}}$}}4_{\dot{d}}5_{\raisebox{-2pt}{${\scriptstyle\dot{e}}$}}1_{\raisebox{-2pt}{${\scriptstyle\dot{a}}$}}\rangle\right.\\
\nonumber &&
~~~~~~
+q_{3}(\displaystyle{\not}p_4 \displaystyle{\not}p_5 \displaystyle{\not}p_1\displaystyle{\not} p_2
-
\displaystyle{\not}p_4\displaystyle{\not} p_2 \displaystyle{\not}p_1 \displaystyle{\not}p_5)
\,\tilde{\lambda}_{4\dot{d}}
\,\langle 5_{\raisebox{-2pt}{${\scriptstyle\dot{e}}$}}1_{\raisebox{-2pt}{${\scriptstyle\dot{a}}$}}2_{\dot{b}}3_{\raisebox{-2pt}{${\scriptstyle\dot{c}}$}}\rangle
\\
\nonumber
&&
~~~~~~\left.+(q_{3}+q_{4})
(\displaystyle{\not}p_5\displaystyle{\not} p_1\displaystyle{\not} p_2\displaystyle{\not} p_3
-
\displaystyle{\not}p_5\displaystyle{\not} p_3\displaystyle{\not}p_2 \displaystyle{\not}p_1)
\,\tilde{\lambda}_{5\dot{e}}
\,\langle1_{\raisebox{-2pt}{${\scriptstyle\dot{a}}$}} 2_{\dot{b}}3_{\raisebox{-2pt}{${\scriptstyle\dot{c}}$}}4_{\dot{d}}\rangle+c.c.\right]\Big\}.~~~~~
\eqae
where $\langle 1_{\raisebox{-2pt}{${\scriptstyle\dot{a}}$}}2_{\dot{b}}3_{\raisebox{-2pt}{${\scriptstyle\dot{c}}$}}4_{\dot{d}}\rangle=\epsilon_{ABCD}\lambda^A_{\raisebox{-2pt}{${\scriptstyle1\dot{a}}$}}\lambda^B_{\raisebox{-2pt}{${\scriptstyle2\dot{b}}$}}\lambda^C_{\raisebox{-2.5pt}{${\scriptstyle3\dot{c}}$}}\lambda^D_{4\dot{d}}$.

\vspace{1mm}
Dimensional reduction of the $\cn=(1,0)$ amplitudes gives the 4d $\mathcal{N}=2$ super Yang-Mills in the $\Phi$-$\Psi$ formalism. We first note that the two six-dimensional superfield corresponds to the half-Fourier transformation of the four-dimensional superfields:
\eq
\Upsilon^{\dot{1}}=-\int d\eta_2e^{\eta_{2}\bar{\eta}^2}\,\Psi_{\mathcal{N}=2}\,,
~~~~~~~~~
\Upsilon^{\dot{2}}=-\int d\eta_2e^{\eta_{2}\bar{\eta}^2}\,\Phi_{\mathcal{N}=2}
\,.
\eqe
The other crucial ingredient is the dimensional reduction of the four-spinor bracket. Since the four-spinor bracket carries four free SU(2) little group indices, and each SU(2) index corresponds to $+\frac{1}{2}$ or $-\frac{1}{2}$ helicity spinors in four dimensions, the bracket contains $2^4=16$ terms in four dimensions. However, most  vanish and the only non-vanishing brackets are
\eqa
\nonumber\langle i_aj_bk_cl_d\rangle~\rightarrow~\langle i_1j_1k_2l_2\rangle=-[ij]\langle kl\rangle\,,
\hspace{9mm}
{}[i_{\dot{a}}j_{\dot{b}}k_{\dot{c}}l_{\dot{d}}]~\rightarrow~[i_{\dot{1}}j_{\dot{1}}k_{\dot{2}}l_{\dot{2}}]=-[ij]\langle kl\rangle\,.
\label{fourspin}
\eqae

As an example, we show that the dimensional reduction of the 6d four-point amplitude gives the 4d $\mathcal{N}=2$ SYM amplitude. We start in 4d  and perform the half-Fourier transform of the $\mathcal{N}=2$ SYM 4-point superamplitude given in eq.~(\ref{MHVgenN}). Using the identity \cite{Hatsuda:2008pm}
\eq
\int \prod_{i=1}^4d\eta_{i2}e^{\eta_{i2}\bar{\eta}_i^2}\left(\sum_{l,k}\eta_{2l}\langle lk\rangle\eta_{2k}\right)=\left(\sum_{1\leq i<j\leq 4}\bar{\eta}^2_{i}[ij]\bar{\eta}^2_{j}\right)\left(\frac{\langle12\rangle\langle23\rangle\langle34\rangle\langle41\rangle}{[12][23][34][41]}\right)^{\frac{1}{4}}\,,
\eqe
we find
\eqa
\nonumber\int \Big( \prod_{i=1}^4d\eta_{i2}e^{\eta_{i2}\bar{\eta}_i^2} \Big)
\,\mathcal{A}_4
&=&\nonumber-\frac{\epsilon_{ijkl}\langle i j\rangle[lk]}{st }\Big(\sum^4_{i=1}q^2_i\Big)\Big(\sum^4_{i=1} \tilde{q}_1\Big)\,,
\eqae
which is exactly the dimensional reduced form of the 6d 4-point amplitude in eq.\,(\ref{6d4ptN1}). The choice of external multiplets $\dot{a}\dot{b} \dot{c} \dot{d}$ as two 1's and two 2's correspond to the choice of which two states are $\Phi$ and which are $\Psi$.

\section{Pure $\mathcal{N}<8$ SG amplitudes}
\label{s:SG}
The formalism for non-maximal SYM can be straightforwardly extended to supergravity amplitudes. Here we outline the procedure for obtaining the $\mathcal{N}<8$ supergravity MHV amplitudes in the $\Phi$-$\Psi$ formalism.

The on-shell states of the $\mathcal{N}<8$ supergravity multiplet can be
neatly packaged into two superfields $\Phi,\Psi$, with the positive helicity graviton appearing as the leading component of $\Phi$, and the negative helicity graviton at the top of $\Psi$. Similar to the discussion of embedding the $\mathcal{N}<4$ SYM multiplets within the maximal one, the two superfields can be obtained from the maximal $\mathcal{N}=8$ superfield $\Phi_{\mathcal{N}=8}$ as
\bea
  \label{sugraN<8}
 \Phi_{\mathcal{N}<8}=\Phi_{\mathcal{N}=8}\bigg|_{\eta_{\mathcal{N}+1},\ldots,\eta_{8}\rightarrow0},\;\;\;
 ~~~~~~~~
 \Psi_{\mathcal{N}<8}=\int \prod_{i=\mathcal{N}+1}^8d\eta_{i}\Phi_{\mathcal{N}=8}\,.
\eea

The MHV amplitude of the $\mathcal{N}=8$ theory can be conveniently written as
 \bea
 \label{MHVsugra}
 \nonumber&&\mathcal{A}_n^{\rm MHV}=\frac{M_{n}(1^- 2^- 3^+ \ldots n^+)}{\langle 12\rangle^8}\delta^{(16)}\left(\sum^n_{i=1}|i\rangle \eta_{ia}\right)\\
&& {\rm with}\;\;\;
~~~~\delta^{(16)}\left(\sum^n_{i=1}|i\rangle \eta_{ia}\right)=\frac{1}{2^8}\prod_{a=1}^8\sum_{i,j}^n\langle ij\rangle \eta_{ia}\eta_{ja} \, ,
 \eea
 where $M_{n}(1^- 2^- 3^+ \ldots n^+)$ is the MHV graviton $n$-point amplitude; at tree level it is unaffected by any other fields. To obtain the MHV superamplitudes for the $\mathcal{N}<8$ theories, we choose the $i$th and $j$th particles to be in the $\Psi$ multiplet and integrate the $8-\mathcal{N}$ $\eta_i$'s and $\eta_j$'s from the $\mathcal{N}=8$ MHV amplitude. Explicitly
\bea
\nonumber \mathcal{F}^{\mathcal{N}}_{n,ij}&=&\left. \int d^{8-\mathcal{N}}\eta_{i}d^{8{\rm-}\mathcal{N}}\eta_{j}
\frac{M_{n}(1^+ \ldots i^- \ldots j^- \ldots n^+)}{\langle ij\rangle^8}
~\delta^{(16)}\left(\sum^n_{k=1}|k\rangle \eta_{ka}\right)\right|_{\eta_{\mathcal{N}+1},\ldots ,\eta_{8}\rightarrow0}\\
&=&(-1)^{\frac{1}{2}\mathcal{N}(\mathcal{N}-1)}
\frac{M_{n}(1^+ \ldots i^- \ldots j^- \ldots n^+)}{\langle 12\rangle^\cn}
~\delta^{(2\mathcal{N})}\left(\sum^n_{k=1}|k\rangle \eta_{ka}\right)
\,.
 \eea
The ${n \choose 2}$ different choices of $i,j$ for $ \mathcal{F}^{\mathcal{N}}_{n,ij}$ are related to each other by simple momentum relabeling.\footnote{This was not the case for the color-ordered SYM amplitudes.} The $\cn=7$ formalism provides an alternative encoding of the  $\cn=8$ supergravity amplitudes; having non-manifest SUSY may be useful in some applications.


\section{Outlook}
\label{sec:disc}

We have presented a uniform approach to $\mathcal{N}<4$ SYM amplitudes in terms of a 4d on-shell superspace formulation that allow us to encode the amplitudes into superamplitudes. The tree superamplitudes are simply truncations of the $\cn=4$ SYM tree superamplitudes, while at loop level one must take into account the different state sums, for example  when evaluating unitarity cuts to reconstruct the loop amplitudes.

A truncation prescription from maximal SUSY to lower SUSY can also be applied in other dimensions. We have demonstrated this in 6 dimensions. An interesting by-product of the 6d analysis is the curious relationship between 4d amplitudes of different helicity structures, i.e.~in different  N$^K$MHV sectors. 
This could be interesting to explore further; one would benefit from knowledge of  compact  explicit expressions for $n\!>\!5$-point tree amplitudes in 6d.

An interesting direction is to explore renormalization using on-shell techniques. On general grounds, we might expect the sum of bubble coefficients, which capture the UV divergences,  to be proportional to a tree amplitude. From  a computational point of view this is not at all obvious, and for N$^K$MHV amplitudes, this becomes even more non-trivial since even the tree-level amplitudes take more complicated form. It will be interesting to explore if there is a simple on-shell mechanism that guarantees the sum of bubble coefficients to be proportional to a tree amplitude.

Another interesting, and phenomenologically relevant, generalization of our work is to couple matter multiplets to the $\cn=1,2$ SYM theories. In that case, it is natural to include a set of Grassmann book-keeping variables for each of the matter multiplets; these can be labeled by the a flavor index $A$.
As   a fairly trivial example, consider $\mathcal{N}=4$ SYM as $\mathcal{N}=1$ SYM coupled to three chiral multiplets with $SU(3)$ flavor symmetry. Or as $\mathcal{N}=2$ SYM coupled to two hypermultiplets with $SU(2)$ flavor symmetry. In both cases, the original $\eta$'s, which are fundamentals of the $SU(4)$ R-symmetry, now transforms under $SU(\mathcal{N})\times SU(4-\mathcal{N})$,
i.e.~$\cn$ $\eta$'s carry R-symmetry index while the $4-\cn$ other ones carry flavor symmetry index. Thus by assigning the original R-index into the reduced R-symmetry index plus flavor indices, one obtains a superamplitude defined on the reduced on-shell superspace plus flavor space. This is of course a trivial rewriting of the $\cn=4$ superamplitudes, but it may give a hint about what to expect for $\cn=1,2$ SYM with matter. The work \cite{Lal:2009gn} together with our results here would be a useful starting point of obtaining explicit results for tree- and loop-level superamplitudes in $\cn=1,2$ SYM with matter.


\section*{Acknowledgements}

We are grateful to Lance Dixon for notes that clarify the  UV/IR properties of the bubble contributions to the 1-loop amplitudes, and to Zvi Bern for further explanations. 
We thank Nima Arkani-Hamed, Dan Freedman, Michael Kiermaier,  David McGady and Jaroslav Trnka for useful discussions.

HE and CP are supported by NSF CAREER Grant PHY-0953232, and in part by the US Department of Energy under DOE grants DE-FG02-95ER 40899. HE was also supported by the Institute for Advanced Study (DOE grant DE-FG02-90ER40542) while this work was in progress. YH is supported by the US Department of Energy under contract DE-FG03-91ER40662.

\appendix

\section{Derivation of NMHV superamplitude for $\cn<4$ SYM}
\label{app:NMHV}

In this appendix we present the derivation of the  tree-level 
NMHV superamplitude formulas for $\cn=1,2,3$ SYM.  For convenience we choose legs $i$ ,$j$ and $n$ to be in the $\Psi$ multiplet; we can always achieved this by using the cyclic symmetry of the color ordered amplitude. One extracts the $\mathcal{N}=1,2,3$ superamplitude via
 \bea
  \cf_{n,(nij)}^{\cn}
 &=&\int d^{4-\cn}\!\eta_i~d^{4-\cn}\!\eta_j~d^{4-\cn}\!\eta_n
 ~\mathcal{A}_n^\text{$\mathcal{N}=4$ NMHV} \,,
 \eea
where  $\mathcal{A}_n^\text{$\mathcal{N}=4$ NMHV}$ was given in \reef{NMHVn4} and $d^{4-\cn}\!\eta_k \equiv \prod_{a=\cn+1}^4d\eta_{ka}$. We rearrange the integration measure and the integrand, keeping careful track of the signs, to find
 \bea
  \cf_{n,(nij)}^{\cn}
 &=&
 (-1)^{\cn}\,\,
 \frac{\delta^{(2\cn)}(\sum |k\>\eta_k)}{\<12\>\cdots \<n1\>}
  \,
 \sum_{2 \le s<t<n-1}
  \bigg(R^{\,\mathcal{N}}_{n;st}~
 \prod_{a=\cn+1}^4 I_{nst,a}\bigg) \, ,
 \label{fromN4}
 \eea
where $R^{\,\mathcal{N}}_{n;st}$ was given in \reef{RN} and
\bea
  I_{nst,a} \equiv \int d\eta_{ia} d\eta_{ja} d\eta_{na}~
  \delta^{(2)} (\sum |k\>\eta_k ) ~
  \bigg(
  \sum_{l=t}^{n-1}\langle n|x_{ns}x_{st}|l\rangle \eta_{la}+\sum_{l=s}^{n-1}\langle n|x_{nt}x_{ts}|l\rangle \eta_{la}
  \bigg)\,.
  \label{Iint}
\eea
Here we have used the explicit form of $\Xi_{nst,a}$ given in \reef{XiDef}.
To carry out the integral \reef{Iint}, one carefully keeps track of the possible orderings of $i,j$ with respect to legs $s,t$. This gives
\begin{itemize}
  \item $i<j <s <t\leq n-1$: The ranges of the two sums in \reef{Iint} do not include  $i$ or $j$, so the integrand is independent of $\eta_{ia}$, $\eta_{ja}$ and $\eta_{na}$, and hence the integral vanishes.
  \item $i<s\leq j<t\leq n-1$: The second term in (\ref{Iint}) contains $\eta^a_j$, so the integral is
  \eq
  I_{nst,a} ~=~\langle i n\rangle \langle n|x_{nt}x_{ts}|j\rangle \, .
  \eqe
  \item $i<s<t\leq j \leq n-1$: Both terms in (\ref{Iint}) contribute to the $\eta_j$ integration. We obtain
  \eqa
I_{nst,a} &=& \langle n i\rangle \langle nj\rangle x^2_{st}\,.
  \eqae
  where we have used the identity $x_{ns}x_{st}+x_{nt}x_{ts}+x^2_{st}=0$.
  \item $2\leq s\leq i<j<t\leq n-1$: The second term in (\ref{Iint}) contains both $\eta_i$ and $\eta_j$, and we use Schouten to find
\eqa
 I_{nst,a}  &=&\langle nj\rangle\langle n|x_{nt}x_{ts}|i\rangle-\langle ni\rangle\langle n|x_{nt}x_{ts}|j\rangle
 =\langle ij\rangle\langle n|x_{nt}x_{ts}|n\rangle \, .
 \eqae
 \item $2\leq s\leq i<t\leq j \leq n-1$: both terms in (\ref{Iint}) contribute and we have
\begin{equation}
I_{nst,a}
 ~=~\langle nj\rangle\langle n|x_{nt}x_{ts}|i\rangle-\langle ni\rangle\langle n|x_{nt}x_{ts}|j\rangle-\langle ni\rangle\langle n|x_{ns}x_{st}|j\rangle
 =\langle j n\rangle\langle n|x_{ns}x_{st}|i\rangle \, .~~~
\end{equation}
\item $2\leq s<t \leq i  < j \le n-1$: By the $\delta^{(2)}$-function, the sums in \reef{Iint} can be converted to run from $n$ to $s-1$ or $t-1$. They are then independent of $\eta_{ia}$, $\eta_{ja}$ and $\eta_{na}$, and hence the integral vanishes.
\end{itemize}

The analysis shows that \reef{fromN4} splits into four sums corresponding to the four orderings with nonzero results above. For each $\cn$, one obtains $\cn$ factors of $I_a$. The rearrangement of the integration measure produces and overall sign. Putting all things together, the result for the $\cn=0,1,2,3,4$ NMHV superamplitude is
\eqa
&&
\hspace{-10mm}
\;\mathcal{F}^\cn_{n,ijn}=
(-1)^{\cn}
\frac{\delta^{(2\cn)}\left(\sum |\lambda\>\eta_\l\right)}{\langle12\rangle\langle23\rangle\cdot\cdot\cdot\langle n1\rangle}\times\\
\nonumber&&
\Bigg[\sum_{i<s\leq j<t\leq n-1}
\Big(\langle in\rangle \langle n|x_{nt}x_{ts}|j\rangle \Big)^{4-\cn}
R^{\,\mathcal{N}}_{nst}
+\sum_{i<s<t\leq j \leq n-1} \Big(\langle ni\rangle\langle nj\rangle x^2_{st} \Big)^{4-\cn}
R^{\,\mathcal{N}}_{nst}
\\
\nonumber
&& +\sum_{2\leq s\leq i<j<t\leq n-1} \Big(\langle ij\rangle\langle n|x_{nt}x_{ts}|n\rangle \Big)^{4-\cn}R^{\,\mathcal{N}}_{nst}
+\sum_{2\leq s\leq i<t\leq j} \Big(\langle jn\rangle\langle n|x_{ns}x_{st}|i\rangle \Big)^{4-\cn}R^{\,\mathcal{N}}_{nst}\Bigg] . 
\eqae
Recognizing the overall factor as $(-1)^{\frac{1}{2}\cn(\cn+1)}\<ij\>^{\cn-4}\mathcal{F}^\cn_{n,ij}$ --- the MHV superamplitude with $\Psi$-sector states $i$ and $j$ --- we arrive at the final result for the NMHV superamplitude \reef{NMHVresN}.

\section{Signs in intermediate state sums}
\label{app:B}

The signs in section \ref{s:sums} are easily verified starting from the equivalent state sum in $\cn=4$ SYM:
\bea
   \nonumber
   &&
   \hspace{-1cm}
   \int d^4\eta_{\ell_1} \,d^4\eta_{\ell_2} ~\ca_L~\ca_R
   ~=~
    -\int  d\eta_{\ell_1,1}\, d\eta_{\ell_2,1}
    \int d^3\eta_{\ell_1}\,d^3\eta_{\ell_2} ~\ca_L~\ca_R\\[2mm]
\label{N4cut}
    &&\to~
    -\int  d\eta_{\ell_1,1}\, d\eta_{\ell_2,1}
    \Big\{
      \ca_L(\dots)~\ca_R(\dots \Psi_{\ell_1} \Psi_{\ell_2})
     +  \ca_L(\dots \Psi_{\ell_1} \Psi_{\ell_2})~\ca_R(\dots ) \\[2mm]
   \nonumber
    &&~  \hspace{3.5cm}
     + \ca_L(\dots \Psi_{\ell_1})~\ca_R(\dots  \Psi_{\ell_2})
     -  \ca_L(\dots  \Psi_{\ell_2})~\ca_R(\dots\Psi_{\ell_1} )
    \Big\} .
\eea
In the second line, we have truncated the state sum into the four possible $\cn=1$ pieces. Next we apply with $\int d^3\eta_i\,d^3\eta_j$ to select $i$ and $j$ to be the external $\Psi$-states (assuming that the 1-loop amplitude is MHV).
If we assume that both subamplitudes at MHV, then there are two cases,
depending on whether $i$ and $j$ sit on the same or on different subamplitudes.

If $i,j\in L$, then only the first term in \reef{N4cut} contributes and we get
\begin{equation}
-\int  d\eta_{\ell_1}\, d\eta_{\ell_2}~
      \ca_L(\dots \Psi_{i} \dots\Psi_{j} )~\ca_R(\dots \Psi_{\ell_1} \Psi_{\ell_2})
 = \int  d\eta_{\ell_1}\, d\eta_{\ell_2}~
      \ca_L(\dots \Psi_{i} \dots\Psi_{j} )~\ca_R(\dots \Psi_{\ell_2}\Psi_{\ell_1}) \, .
\end{equation}
We used cyclic symmetry and that the $\Psi$'s are Grassmann odd. We recognize Cut (a) from section \reef{s:sums}.

If $i \in L$ and $j \in R$, the last two terms of \reef{N4cut} contribute and this gives
\begin{equation}
 \int  d\eta_{\ell_1}\, d\eta_{\ell_2}~
          \Big\{\ca_L(\dots \Psi_{i} \dots\Psi_{\ell_1})~\ca_R(\dots  \Psi_{j} \dots\Psi_{\ell_2})
     -  \ca_L(\dots \Psi_{i}\dots  \Psi_{\ell_2})~\ca_R(\dots\Psi_{j} \dots  \Psi_{\ell_1} ) \Big\}\,.
\end{equation}
This is Cut (b) of section \reef{s:sums}, as can be seen after rearrangement of the $\Psi$'s in $\ca_R$.

The signs in the $\cn=2$ and $\cn=3$ intermediate state sums can be checked similarly.

\vspace{5mm}


 \end{document}